\documentclass[journal=jacsat,manuscript=article]{achemso}

\usepackage[version=3]{mhchem} % Formula subscripts using \ce{}

\usepackage{comment}
\usepackage{xcolor}
\usepackage{graphicx}
\usepackage{braket}
\usepackage{amsmath}
\usepackage{subcaption} % Required for creating subfigures
\definecolor{rygreen}{rgb}{1.0, 0.5, 0.0}
\newcommand{\rz}[2]{{\color{rygreen}{#1}}}
\definecolor{yka}{rgb}{1.0, 0.0, 0.0}
\newcommand{\yk}[2]{{\color{yka}{#1}}}

\definecolor{vbcolor}{rgb}{1.0, 0.7, 0.3}

\usepackage{amssymb}

\bibstyle{achemso}

\title[An \textsf{achemso} demo]
  {All-electron BSE@GW method  with Numeric Atom-Centered Orbitals for Extended Systems}

\author{Ruiyi Zhou}
\affiliation{Department of Chemistry, University of North Carolina at Chapel Hill, Chapel Hill, North Carolina 27599, USA}
\author{Yi Yao}
\affiliation{Department of Chemistry, University of North Carolina at Chapel Hill, Chapel Hill, North Carolina 27599, USA}
\alsoaffiliation{Thomas Lord Department of Mechanical Engineering and Materials Science, Duke University, Durham, North Carolina 27708, USA}

\author{Volker Blum}
\affiliation{Thomas Lord Department of Mechanical Engineering and Materials Science, Duke University, Durham, North Carolina 27708, USA}
\alsoaffiliation{Department of Chemistry,  Duke University, Durham, North Carolina 27708, USA}

\author{Xinguo Ren}
\affiliation{Institute of Physics, Chinese Academy of Sciences, Beijing 100190, China}
\email{renxg@iphy.ac.cn}

\author{Yosuke Kanai}
\affiliation{Department of Chemistry, University of North Carolina at Chapel Hill, Chapel Hill, North Carolina 27599, USA}
\alsoaffiliation{Department of Physics and Astronomy, University of North Carolina at Chapel Hill, Chapel Hill, North Carolina 27599, USA}

\email{ykanai@unc.edu}

\begin{document}

\begin{abstract}
Green's function theory has emerged as a powerful many-body approach not only in condensed matter physics but also in quantum chemistry in recent years. We have developed a new all-electron implementation of the BSE@GW formalism using numeric atom-centered orbital basis sets (Liu et al., J. Chem. Phys. 152, 044105 (2020)). We present our recent developments in implementing this formalism for extended systems with periodic boundary conditions. We discuss its numerical implementation and various convergence tests pertaining to numerical atom-centered orbitals, auxiliary basis sets for the resolution-of-identity formalism, and Brillouin zone sampling. Several proof-of-principle examples are presented to compare with other formalisms, illustrating the new all-electron BSE@$GW$ method for extended systems.
\end{abstract}

\section{1.Introduction }
In the last decade, the many-body perturbation theory based on Green's function formalism has found its way into chemistry community from the condensed matter physics community. By going beyond well-accepted approximations for condensed matter systems (e.g. plasmon-pole approximation, etc), a number of groups have shown that the so-called $GW$ and Bethe-Salpeter equation (BSE) methods can be made quite promising for studying excited state properties of isolated molecules \cite{van2013gw,leng2016gw,golze2019gw, blase2020bethe}.
Solving the particle-hole two-particle Green's function, the BSE method \cite{strinati1984effects,albrecht1998ab,rohlfing1998electron,onida2002electronic} has become increasingly popular for calculating neutral excitations of molecules in recent years \cite{blase2020bethe}, adding to a history of successes for condensed matter systems\cite{marsili2017large,vorwerk2019bethe,deslippe2012berkeleygw}. It is now widely recognized as a promising alternative\cite{faber2014excited,jacquemin2017bethe, blase2018bethe} to density functional theory (DFT)-based approaches such as linear-response time-dependent density function theory (LR-TDDFT)\cite{casida1995time,ullrich2011time} and traditional wavefunction-based methods like the equation-of-motion coupled cluster (EOM-CC) \cite{bartlett2012coupled,krylov2008equation}. 

The BSE method is based on the many-body perturbation theory in the Green's function ($G$) framework, and the screened Coulomb interaction, $W$, is used to model the interaction between the excited electron and hole. Combined with the $GW$ approximation to the self-energy, the BSE method has been shown to successfully yield the optical spectra of solids and nano-structured systems, and more recent work also shows similar applicability to low-energy electronic excitation of molecular systems \cite{yang2007excitonic,perebeinos2005radiative,jacquemin2015benchmarking,jacquemin2016assessment,körbel2014benchmark}. The BSE@$GW$ approach yields the accuracy of 0.1-0.2 eV, being comparable to the EOM-CCSD \cite{jacquemin2017bethe,blase2018bethe}. 
Due to its favorable scaling of $N^4$ in terms of the number of electrons, the approach is highly promising for studying excited-state properties of increasingly complex systems. Our previous work has also shown that all-electron BSE@$GW$ method using atom-centered orbital basis sets  provides high accuracy for modeling core-electron excitations, comparable to the state-of-the-art EOM-CCSD method\cite{yao2022all}. 
This computational capability to quantitatively predict X-ray absorption spectra of molecules is of great interest as many light-source facilities have undergone great advancement in recent years.

Building on our recent effort on developing \textit{all-electron} many-body perturbation theory methods using numeric atom-centered orbital (NAO) basis sets (e.g., $GW$ methods for extended periodic systems \cite{ren2021all} and the BSE for isolated systems \cite{liu2020all,yao2022all}), we here extend the BSE@$GW$ approach to periodic systems with Brillouin zone (BZ) integration. For studying  extended systems, taking into account the dependence on the reciprocal wave vector requires careful consideration.
Most $GW$/BSE method developments originating in condensed matter physics are based on plane-waves or real-space-grids with non-local pseudo-potentials, and the numerical formulations of these Green's function methods are largely incompatible with the mathematical/numerical frameworks used in many quantum-chemistry developments/codes. 
The present theoretical method and algorithm will benefit the quantum chemistry field, which is largely based on all-electron implementations with atom-centered basis functions including Gaussian-type orbitals. Furthermore, some of us have recently demonstrated significant efficiency increases of numerically precise exact exchange and hybrid DFT for periodic system sizes exceeding 10,000 atoms in size, using the NAO formalism.\cite{Kokott2024} Thus, the ground work laid in the present paper for BSE@$GW$ should be extendable to significantly larger scales as well in future developments.
Overall, our development will enable the quantum chemistry community to take advantage of recent exciting  advances in the methods based on Green's function theory and promote further synergies with traditional post-Hartree-Fock methods. 

%%%%%%%%%%%%%%%%%%%%%%%%%%%%%%
\section{2. Green's Function Theory}

\subsection{2.1 Bethe Salpeter Equation}

For describing electron-hole pairs in a many-electron system, 
the two-body correlation function
%, denoted here as $ {\cal L}(12;1^{\prime}2^{\prime})$ %(occasionally represented as $\cal L$ in some literature \yk{we should probably use $\cal L$ - it's customary as you do in Eq. 12} : see https://www.cond-mat.de/events/correl16/manuscripts/reining.pdf) 
plays a central role in many-body perturbation theory based on Green's function. 
One-body Green's function $G_1$ and two-body Green's function $G_2$ are defined as 
\begin{equation}\label{eq:G1}
    iG_1(1,2)=\langle N,0|T[\hat{\psi}(1)\hat{\psi}^\dagger(2)]|N,0\rangle
\end{equation}
\begin{equation}\label{eq:G2}
    i^2G_2(1,2;1^{\prime},2^{\prime})=\langle N,0|T[\hat{\psi}(1)[\hat{\psi}(2)\hat{\psi}^\dagger(2^{\prime} )\hat{\psi}^\dagger(1^{\prime})]|N,0\rangle
\end{equation}
where the creation operator $\hat{\psi}(1)$ and the annihilation operator $\hat{\psi}^\dagger(2)$ are the field operators written in the Heisenberg picture: $\hat{\psi}(1) = e^{i\hat{H}t_1}\hat{\psi}(\mathbf{x_1)}e^{-i\hat{H}t_1}$ and $\hat{\psi}^{\dagger}(2) = e^{i\hat{H}t_2}\hat{\psi}^{\dagger} (\mathbf{x_2})e^{-i\hat{H}t_2}$, $1 \equiv (x_1, t_1)$, $2 \equiv (x_2, t_2)$ denotes a composite variable encompassing space, spin, and time.  Here, $|N,0 \rangle$ is the ground state state for a N-electron system and $T$ is the time ordering operator.
%definded as:
%\begin{equation}
%T[A(t_1),B(t_2)] \equiv \begin{cases} A(t_1)B(t_2) ,&  t_1 > t_2 \\   B(t_2)A(t_1) , & t_1 <t_2 \end{cases}
%\end{equation}}
The correlation function, $\cal L$, is formally given as a functional derivative of the one-body Green's function $G(1, 1^{\prime})$ with respect to an external non-local perturbation $U(2^{\prime},2)$,
\begin{equation}
\label{eq:L_derivative_def}
    {\cal L}(1,2;1^{\prime},2^{\prime})= \frac{\delta G(1, 1^{\prime})}{\delta U(2^{\prime},2)}
\end{equation}
In terms of the Greens' functions, it is expressed as 
\begin{equation}\label{eq:L_def}
     {\cal L}(1,2;1^{\prime},2^{\prime})=-G_2(1,2;1^{\prime},2^{\prime})+ G_1(1,1^{\prime})G_1(2,2^{\prime})
\end{equation}
${\cal L}(1,2;1^{\prime},2^{\prime})$  describes the probability amplitude of an electron propagating from $1^{\prime}$ to $2$ and a hole propagating from $1$ to $2^{\prime}$. 
By incorporating Eqn. \ref{eq:G1} and Eqn. \ref{eq:G2} into Eqn. \ref{eq:L_def} and using the completeness property of N-electron system in an excited state $S$ $|N,S\rangle$ (i.e. $\sum_{S}|N,S\rangle \langle N,S|=I$), we can expand the correlation function as following:
%within the time domain \yk{time domain? $\tau$ is probably the time difference here},
\begin{equation}\label{eq:chi_time_domain}
\begin{aligned}
        {\cal L}(\mathbf{x_1}, \mathbf{x_2}; \mathbf{x_1^{\prime}}, \mathbf{x_2^{\prime}}; \tau)&= \theta(\tau)\sum_{S \neq 0 }\langle N,0|\hat{\psi}^\dagger (\mathbf{x_1^{\prime}})\hat{\psi} (\mathbf{x_1})|N,S\rangle \langle N,S|\hat{\psi}^\dagger (\mathbf{x_2^{\prime}})\hat{\psi} (\mathbf{x_2})|N,0 \rangle  e^{-i(E_{N,S}-E_{N,0})\tau}\\& +
        \theta(-\tau)\sum_{S \neq 0 }\langle N,0|\hat{\psi}^\dagger (\mathbf{x_2^{\prime}})\hat{\psi} (\mathbf{x_2})|N,S\rangle \langle N,S|\hat{\psi}^\dagger (\mathbf{x_1^{\prime}})\hat{\psi} (\mathbf{x_1})|N,0\rangle e^{i(E_{N,S}-E_{N,0})\tau}
\end{aligned}
\end{equation}
where $E_{N,S}$ is the energy for $|N,S\rangle$. Due to time translation invariance, the linear-response function depends only on $\tau = t_1-t_2$.
By employing Fourier transformation, we arrive at the Lehmann representation of the two-body correlation function $\cal L$  within the frequency domain
%\xr{assuming $t_1 = t'_1, t_2=t_2'$, and $\tau = t'_1 = t_1?$}
\begin{equation}\label{eq:chi_lehmann}
{\cal L}\left(\mathbf{x_1}, \mathbf{x_2}; \mathbf{x_1^{\prime}}, \mathbf{x_2^{\prime}} ; \omega\right)=i\sum_{S \neq 0}[\frac{{A_S(\mathbf{x_1},\mathbf{x_1^{\prime}})}A^*_S(\mathbf{x_2^{\prime},\mathbf{x_2})}}{\omega-\omega_S+i 0^{+}} -\frac{{A_S(\mathbf{x_2},\mathbf{x_2^{\prime}})}A^*_S(\mathbf{x_1}^{\prime},\mathbf{x_1})}{\omega+\omega_S-i 0^{+}}]
\end{equation}
%\xr{$A_n$ in the numerator should $A_m$.}
where  $0^+$ is a positive infinitesimal, $A_S$ represents the amplitude associated with the coupled electron-hole pair for the $S$-th excited state:$A_S(\mathbf{x},\mathbf{x^{\prime}})=-\langle N,0 |\hat{\psi}^{\dagger}(\mathbf{x})\hat{\psi}(\mathbf{x}^{\prime})|N,S\rangle $ 
%\yk{|N> to be defined?} 
and $\omega_m$ corresponds to the excitation energy from the ground state to the $S$-th excited state $\omega_S=E_{N,S}-E_{N,0}$.

%%%%%%%%% BSE 

The Bethe-Salpeter equation (BSE) relates the two-body correlation function to the non-interacting one as\cite{nakanishi1969general}
\begin{equation}\label{eq:bse general}
    {\cal L}(1,2;1^{\prime},2^{\prime})={\cal L}_{0}(1,2;1^{\prime},2^{\prime})+\int d(3456){\cal L}_{0}(1,4;1^{\prime},3)K(3,5;4,6){\cal L}(6,2;5,2^{\prime})
\end{equation}
where ${\cal L}_0(1,2;1^{\prime},2^{\prime} )$ is non-interacting correlation function, given by 
\begin{equation}\label{eq:chi_0_def}
     {\cal L}_{0}(1,2;1^{\prime},2^{\prime})=G_1(1,2^{\prime})G_1(2,1^{\prime}).   
\end{equation}
The electron-hole interaction kernel, $K(35;46)$, is defined as

\begin{equation}
    K(3,4;5,6)=\frac{\delta[ v_{H}(3)\delta(3,4)+\Sigma(3,4)]}{\delta G(6,5)}
\end{equation}
where $ v_{H}(1)=-i\int d2 v(1,2)G(2,2^+)$ is the Hartree potential and  $\Sigma(3,4)$ is the self-energy.   $1^+$ is used to indicates $t_1^+ = t_1 + 0^+$.
By adopting the widely-used $GW$ approximation\cite{hedin1965new} to the self-energy $\Sigma(3,4)= i G(3,4)W(3,4)$, the kernel simplifies to\cite{rohlfing2000electron},
\begin{equation}\label{eq:K}
    K(3,4;5,6)=-i \delta(3,4)\delta(5^-,6)v(3,6)+i\delta (3,6)\delta(4,5)W(3^+,4;\omega)
\end{equation}
where $\delta$ represents Dirac delta function. Although the frequency dependence of $W$ has been considered in solving BSE\cite{rohlfing2000electron,loos2020dynamical,zhang2023effect}, most standard implementations adapt the so-called static screening effect approximation by neglecting the frequency dependence \cite{rohlfing2000electron} such that Eq. \ref{eq:K} reduces to
\begin{equation}     K(\textbf{r}_{3},\textbf{r}_{5},\textbf{r}_{4},\textbf{r}_{6})=-i v(\textbf{r}_{3}-\textbf{r}_{5})\delta(\textbf{r}_{3}-\textbf{r}_{4})\delta(\textbf{r}_{5}-\textbf{r}_{6})+i W(\textbf{r}_{3},\textbf{r}_{4})\delta(\textbf{r}_{3}-\textbf{r}_{6})\delta(\textbf{r}_{4}-\textbf{r}_{5})
\end{equation}
In our subsequent discussion, we use the notation $W(\textbf{r},\textbf{r}^{\prime})$ to represent the static screened interaction in the limit of $\omega=0$ for brevity. The BSE interaction kernel includes two physically distinct terms: first, the exchange term, which results from the bare Coulomb potential $v$, and second, the direct interaction term coming from the screened exchange interaction $W$. Note that this is in contrast with the one-electron self-energy case, and can be seen from the structure of Feynman diagrams. The direct interaction term is responsible for the attractive nature of electron-hole interaction and formulation of bound electron-hole states (exciton states). On the other hand, the exchange interaction term controls details of the excitation spectrum, such as the splitting between spin-singlet and spin-triplet excitations. \cite{rohlfing1998electron}. 
%\xr{One needs to be cautious here: when it comes to the two-particle process, the first bare Coulomb terms corresponds to an "exchange process", while the second terms corresponds to an direct interaction terms with screened Coulomb interactions. }
In the limit in which $W$ approaches the value of $v$ as the dielectric term $\epsilon$ approaches one, it reduces to time-dependent Hartree-Fock theory.

%Linking it to bare exchange in Hartree-Fock could cause confusion, because this term is usually called ``direct interaction" in the literature.

\subsection{2.2 Bethe Salpeter Equation as  Eigenvalue Problem}
For first-principles theory implementation of BSE, it is convenient to write the BSE amplitudes in Eq. \ref{eq:chi_lehmann} in terms of the particle-hole basis,

\begin{equation}\label{eq:bse_basis}
A_S(\mathbf{x},\mathbf{x^{\prime}}) = \sum_{vc} X_{vc,S}\psi_c(\mathbf{x})\psi_v^*(\mathbf{x^{\prime}}) + Y_{cv,S}\psi_v(\mathbf{x})\psi_c^*(\mathbf{x^{\prime}})
\end{equation}
%\xr{In the usual convention, there is a complex conjugate $\psi_v^\ast(\mathbf{x^{\prime}}$ in the first term and $\psi_c^\ast(\mathbf{x^{\prime}}$ in the second term. }
where $\psi_c$ and $\psi_v$ represent single-particle orbitals of the conduction band (i.e. unoccupied) and valence band (i.e. occupied), respectively. 
$\psi_c(\mathbf{x})\psi_v^*(\mathbf{x^{\prime}})$ and $\psi_v(\mathbf{x})\psi_c^*(\mathbf{x^{\prime}})$ represent the particle-hole basis functions, and they correspond to resonant and anti-resonant transitions of electron-hole pairs, respectively. Typically, by adapting the $G_0W_0$ approximation for the self-energy calculation, the orbitals from mean-field theories such as Hartree-Fock (HF) and Kohn-Sham density functional theory (KS-DFT) are used for the single-particle orbitals in practice. Thus, the matrices $X$ and $Y$ represent the solutions to the BSE, which needs to be solved. 

The non-interacting correlation function ${\cal L}_0$ (Eq.  \ref{eq:chi_0_def}) can be written using the Lehmann representation, analogously to the derivation of Eq. \ref{eq:chi_lehmann} from Eq. \ref{eq:chi_time_domain},
\begin{equation}\label{eq:chi_0}
{\cal L}_0\left(\mathbf{x_1}, \mathbf{x_2}; \mathbf{x_1^{\prime}}, \mathbf{x_2^{\prime}} ; \omega\right)=  i \sum_{v, c}\left[\frac{\psi_c\left(\mathbf{x}_1\right) \psi_v\left(\mathbf{x}_2\right) \psi_v^*\left(\mathbf{x}_1^{\prime}\right) \psi_c^*\left(\mathbf{x}_2^{\prime}\right)}{\omega-\left(\epsilon_c^{QP}-\epsilon_v^{QP}\right)+i0^+}\right. \left.-\frac{\psi_v\left(\mathbf{x}_1\right) 
\psi_c\left(\mathbf{x}_2\right) \psi_c^*\left(\mathbf{x}_1^{\prime}\right) \psi_v^*\left(\mathbf{x}_2^{\prime}\right)}{\omega+\left(\epsilon_c^{QP}-\epsilon_v^{QP}\right)-i0^+}\right]
\end{equation}
where $\epsilon_n^{QP}$ is the quasi-particle energy for the quais-particle orbital indexed by $n$.

For convenience, we introduce the variable $z=\omega+i0^+$sgn$(f_{n_1}-f_{n_2})$, where $f_n$ is  the occupation number for orbital index $n$. 
In terms of the particle-hole basis (see Eq.  \ref{eq:bse_basis}), we can arrive at the numerically convenient matrix representation for ${\cal L}_0$

\begin{equation}
{\cal L}_0\left(\mathbf{x_1}, \mathbf{x_2}; \mathbf{x_1^{\prime}}, \mathbf{x_2^{\prime}}  ; \omega\right)  =({\cal L}_0)_{n_1n_2;n_3n_4}(z) \psi_{n_1}(\mathbf{x}_1)\psi_{n_2}(\mathbf{x}_2)\psi^*_{n_3}(\mathbf{x}_1^{\prime})\psi_{n_4}^*(\mathbf{x}_2^{\prime})
\end{equation}
where the  diagonal matrix ${\cal L}_0$ is defined as 
\begin{equation}\label{eq:chi_0_basis}
({\cal L}_0)_{n_1n_2;n_3n_4}(z) = i\frac{f_{n_2}-f_{n_1}}{z-(\epsilon_
{n_1}^{QP}-\epsilon_{n_2}^{QP})}\delta_{n_1 n_4}\delta_{n_2 n_3}.
\end{equation}
%\xr{The second term in the numerator is $(1-f_{n_1})f_{n_2}$? If so, then it reduces to $f_{n_2}- f_{n_1}$.}
Noting ${\cal L}^{-1}(z)={\cal L}_0^{-1}(z)-K$, the BSE (Eq. \ref{eq:bse general}) is expressed in the matrix representation as

\begin{equation}\label{eq:chi_matrix_rep}
    ({\cal L})_{n_1n_2;n_3n_4}(z) = [{\cal L}_0(z) - K]^{-1}_{n_1n_2;n_3n_4} = i[H^{2p} - \mathbb{I} z]^{-1}_{n_1n_2;n_3n_4} (f_{n_2} - f_{n_4})
\end{equation}
where $\mathbb{I}$ is the identity matrix, and $H^{2p}$ is the two-particle Hamiltonian\cite{martin2016interacting} given by

\begin{equation}
    (H^{2p})_{n_1n_2;n_3n_4}=(\epsilon_{n_2}^{QP}-\epsilon_{n_1}^{QP})\delta_{n_1n_4}\delta_{n_2n_3}+(f_{n_1}-f_{n_3})(\alpha^{s/t} V-W)_{n_1n_2;n_3n_4}.
\end{equation}
%\xr{This means that the second term may change sign depending $n_1$ and $n_3$ are valence or conduction orbitals?}
Here the spin degree of freedom has been integrated out. 
For transitions with different spins, we have $\alpha^s=2$ for singlet excitations and $\alpha^t=0$ for triplet excitations. The explicit forms of the matrix elements of $V$ and $W$ will be discussed in a later subsection.
Meanwhile,we can observe that  the eigenvalues of $H^{2p}$ in Eq. \ref{eq:chi_matrix_rep} correspond precisely to the poles in the Lehmann representation of ${\cal L}$ (see Eq. \ref{eq:chi_lehmann}). 
The eigenvalues represent the excitation energies, $\omega_S=E_{N,S}-E_{N,0}$, of the $N$-electron system in terms of the particle-hole excitation.
The corresponding eigenvectors give the amplitudes in the BSE as defined in Eq. \ref{eq:bse_basis}.
To simplify the eigenvalue problem of $H^{2p}$, we can separate the particle-hole basis pair $(n_1,n_2)$ into resonant pairs $(i,a)$ and anti-resonant pairs $(a,i)$,  the particle-hole basis pair $(n_3,n_4)$ into $(j,b)$ and $(b,j)$, where the orbital indices $i, j$ refer to the valence band states (i.e. occupied orbitals) while $a, b$ refer to the conduction band states (i.e. unoccupied orbitals). As discussed by Rohlfing et al. \cite{rohlfing2000electron} and Strinati et al. \cite{strinati1984effects}, the BSE is finally written in a numerically convenient form as an eigenvalue equation as often encountered in quantum chemistry and condensed matter physics,
    \begin{gather} \label{eq:BSE_eigenvalue}
 \begin{bmatrix} A & B \\ B^* & A^* \end{bmatrix}
 \begin{bmatrix} X_m \\  Y_m \end{bmatrix}
 =
 \omega_S
  \begin{bmatrix} \mathbb{I} & 0 \\ 0 & -\mathbb{I} \end{bmatrix}
\begin{bmatrix} X_S \\  Y_S \end{bmatrix}
\end{gather}
%\xr{The diagonal matrix (1, 0; 0 -1) should be multiplied either from the left side or right side, but not both. }
Here $\omega_S$ is the excitation energy (see  Eq. \ref{eq:chi_lehmann}), and $(X_S, Y_S)$ correspond to the eigenvectors defined in Eq. \ref{eq:bse_basis}.
The matrix blocks denoted by A correspond to the Hamiltonian describing resonant transitions from occupied to unoccupied orbitals, while the matrix blocks represented by $-A^*$ correspond to the Hamiltonian for anti-resonant transitions from unoccupied to occupied orbitals. Similarly, the blocks represented by B and $-B^*$ account for the coupling between resonant and anti-resonant pairs.
The matrices $A$ and $B$ are given by
\begin{equation} 
A_{ia}^{jb}  = (\epsilon_{a}^{QP} - \epsilon_{i }^{QP})\delta_{ij}\delta_{ab} + \alpha^{S/T} \braket{i a|\hat{V}|j b} - \braket{ij|\hat{W}|ab}
\end{equation}
\begin{equation} 
B_{ia}^{bj}  =   \alpha^{S/T} \braket{i a|\hat{V}|bj}- \braket{i b|\hat{W}|aj}
\end{equation}
The operators $\hat{V}$ and $\hat{W}$ represent the bare Coulomb operator and static screened Coulomb operator, respectively, and numerical evaluation of these matrices are  discussed in the subsequent sections.
This BSE eigenvalue problem has a mathematically similar form as the widely-known Casida equation of LR-TDDFT.\cite{casida1995time} while their theoretical origins are quite different as discussed above. 

Although the blocks $\textbf{A}$ and $-\textbf{A}^*$ 
are Hermitian, the overall matrix is non-Hermitian. Numerically, solving this non-Hermitian eigenvalue problem is quite complicated even though a wide range of efficient eigensolvers have been developed for this specific type of numerical problem in recent years\cite{shao2016bsepack,benner2017fast,benner2018improving,ljungberg2015cubic}.
In practical implementation, the Tamm-Dancoff approximation (TDA) is widely used. The TDA amounts to neglecting the coupling matrix between excitation and de-excitation pairs (i.e. $B$ and $-B^{*}$), reducing the problem to a Hermitian eigenvalue problem
\begin{equation}\label{eq:BSE_eigenvalue_TDA}
    A X_{S}=\omega_SX_{S}
\end{equation}
This simplification is justifiable as long as the energy associated with particle-hole interaction remains significantly smaller than the quasi-particle energy gap. 
As discussed in a previous study \cite{rocca2014ab,ljungberg2015cubic,sander2015beyond,blase2020bethe}, the TDA works particularly well especially for solids in the optical limit. In quantum chemistry, the  TDA is also used in the context of Casida's equation of  LR-TDDFT and the Configuration Interaction Singles (CIS) methods\cite{hirata1999time,chantzis2013tamm}.

%%%%%%%%% extended systems
For extended systems, 
it is straightforward to extend this formalism to the BZ with Bloch states. 
Particle-hole pairs here include those excitations from the valence band at the k-point $\mathbf{k}_1$ to the conduction band at the k-point $\mathbf{k}_1+\mathbf{k}^0$ such that $\mathbf{k}^0$ represents the momentum change in the excitation. Eq. \ref{eq:BSE_eigenvalue} then becomes
\begin{equation} 
\begin{split}
A_{ia\mathbf{k}_1}^{jb\mathbf{k}_2} & = (\epsilon_{a\mathbf{k}_1+\mathbf{k}^0}^{QP} - \epsilon_{i \mathbf{k}_1}^{QP})\delta_{ij}\delta_{ab}\delta_{\mathbf{k}_1\mathbf{k}_2} \\
 & + \alpha^{S/T} \braket{i\mathbf{k_1} a\mathbf{k}_1+\mathbf{k}^0|\hat{V}|j\mathbf{k}_2 b\mathbf{k}_2+\mathbf{k}^0} \\
 & - \braket{i\mathbf{k}_1 j\mathbf{k}_2|\hat{W}|a\mathbf{k}_1+\mathbf{k}^0 b\mathbf{k}_2+\mathbf{k}^0}
\end{split}
\end{equation}
%\begin{equation} \label{eq_B}
%\begin{split}
%B_{ia\mathbf{k_1}}^{jb\mathbf{k_2}} & =  - \alpha^{S/T} \braket{i\mathbf{k_1} a\mathbf{k_1}\mathbf{+k^0}|\hat{V}|b\mathbf{k_2}\mathbf{+k^0} n\mathbf{k_2}} \\
% & + \braket{i\mathbf{k_1} b\mathbf{k_2}\mathbf{+k^0}|\hat{W}|a\mathbf{k_1}\mathbf{+k^0} j\mathbf{k_2}}
%\end{split}
%\end{equation}
In a typical calculation of the optical absorption spectrum, where electron-phonon coupling can be neglected, it is assumed that there is no momentum change involved. Therefore, $\mathbf{k}^0$ vector is generally taken to be zero. 
With this consideration, we construct the Bethe-Salpeter equation (BSE) Hamiltonian and solve the associated Hermitian eigenvalue problem of matrix A in Eq. \ref{eq:BSE_eigenvalue_TDA}.
with
\begin{equation} \label{eq:eq_A}
\begin{split}
A_{ia\mathbf{k_1}}^{jb\mathbf{k_2}} & = (\epsilon_{a\mathbf{k_1}}^{QP} - \epsilon_{i \mathbf{k_1}}^{QP})\delta_{ij}\delta_{ab}\delta_{\mathbf{k_1}\mathbf{k_2}} + \alpha^{S/T} \braket{i\mathbf{k_1} a\mathbf{k_1}|\hat{V}|j\mathbf{k_2} b\mathbf{k_2}}- \braket{ i\mathbf{k_1} j\mathbf{k_2}|\hat{W}|a\mathbf{k_1}b\mathbf{k_2}}
\end{split}
\end{equation}
\begin{equation}\label{eq:bse_kernel_V}
      \begin{split}
        \braket{i\mathbf{k_1} a\mathbf{k_1}|\hat{V}|j\mathbf{k_2} b\mathbf{k_2}} = &\iint d\mathbf{r} d\mathbf{r'} \psi_{i}^\mathbf{k_1}(\mathbf{r}) \psi_{a}^{\mathbf{k_1}*}(\mathbf{r})v(\mathbf{r},\mathbf{r'}) \psi_{j}^{\mathbf{k_2}*}(\mathbf{r'}) \psi_{b}^{\mathbf{k_2}}(\mathbf{r'}) \\ 
        %=&\iint d\mathbf{r} d\mathbf{r'} \psi_{i,\mathbf{k_1}}(\mathbf{r}) \psi_{a,\mathbf{k_1}}^*(\mathbf{r})\frac{1}{|\mathbf{r}-\mathbf{r}'|} \psi_{j,\mathbf{k_2}}^*(\mathbf{r'}) \psi_{b,\mathbf{k_2}}(\mathbf{r'})
      \end{split}
\end{equation}
%\begin{equation} 
%\begin{split}
%&\langle i\mathbf{k_1} j\mathbf{k_2}|\hat{W}|a\mathbf{k_1} b\mathbf{k_2}\rangle
%=  \iint d\mathbf{r} d\mathbf{r'} \psi_{i,\mathbf{k_1}}(\mathbf{r}) \psi_{j,\mathbf{k_2}}^*(\mathbf{r}) \frac{i}{2\pi}\int d\omega'e^{-i\omega \tau } W(\mathbf{r},\mathbf{r'},\omega')\\
%& \left(\frac{1}{\Omega^S-\hbar\omega'-(\epsilon_a^{QP}-\epsilon_i^{QP})+i\tau}+\frac{1}{\Omega^S+\hbar\omega'-(\epsilon_a^{QP}-\epsilon_i^{QP})+i\tau}\right)\psi_{a,\mathbf{k_1}}^*(\mathbf{r'}) \psi_{b,\mathbf{k_2}}(\mathbf{r'})
%\end{split} 
%\end{equation}
\begin{equation}\label{eq:BSE_kernel_W}
\begin{split}
     \langle i\mathbf{k_1} j\mathbf{k_2}|\hat{W}|a\mathbf{k_1} b\mathbf{k_2}\rangle 
     = \iint d\mathbf{r} d\mathbf{r'} \psi_{i}^{\mathbf{k_1}}(\mathbf{r}) \psi_{j}^{\mathbf{k_2}*}(\mathbf{r})W(\mathbf{r},\mathbf{r'}) \psi_{a}^{\mathbf{k_1}*}(\mathbf{r'}) \psi_{b}^{\mathbf{k_2}}(\mathbf{r'})    
\end{split}
\end{equation}
The absorption spectrum, given by $\epsilon_2$, can be obtained from the excitation energy $\omega_S$ and the exciton wave-function coefficients $X_S$ as
\begin{equation}\label{eq:absorption_spectrum_eh}
    \epsilon_2(\omega)=\frac{16\pi^2e^2}{\omega^2}\sum_{S}|\mathbf{e}\cdot \langle 0|\hat{\mathbf{v}}|S\rangle|^2\delta(\omega-\omega_S)
\end{equation}
where 
\begin{equation}
    \langle 0 |\hat{\mathbf{v}}|S\rangle=\sum_{vc\mathbf{k} }\langle v\mathbf{k}|\hat{\mathbf{v}}|c\mathbf{k}\rangle X_{vc\mathbf{k} ,S}
\end{equation}
and $\hat{\mathbf{v}}$ is the velocity operator and $\mathbf{e}$ is the direction of the polarization of light.

First-principles computational methods based on Green's function theory like $GW$ and BSE originate formally from application of quantum field theory in condensed matter physics, and they have traditionally been formulated with plane waves as the basis sets along with the use of pseudo-potentials\cite{rocca2012solution,sander2015beyond,albrecht1998ab,rohlfing1998electron,rohlfing2000electron}.
In recent years, there has been a growing interest in formulating $GW$  and BSE methods using atom-centered basis sets in the context of traditional molecular quantum chemistry \cite{blase2011first,faber2012electron,wilhelm2016gw,wilhelm2017periodic,bruneval2016molgw,zhu2021all,lei2022gaussian,ren2012resolution,ren2021all,golze2019gw,bruneval2015systematic,blase2020bethe,liu2020all,yao2022all}.
Gaussian\cite{zhu2021all} and NAO-based\cite{ren2021all} $GW$ methods have been also demonstrated for extended periodic systems, with reciprocal space summations over the BZ, in recent years. 
As an important prerequisite, we emphasize that the valence and low-lying conduction energy band structures at the Kohn-Sham level of theory are numerically completely converged using ``\textit{tier} 2'' NAO basis sets. In a broad benchmark of DFT-KS PBE band structures between FHI-aims code and another all-electron Wien2k code (full-potential (linearized) augmented plane-wave ((L)APW) + local orbitals (lo) method), average deviations are shown to be $<$0.01~eV and $<$0.02~eV for the valence band range and for the conduction band range up to 5~eV above the conduction band minimum, respectively.\cite{Huhn2017}
In Green's function theory calculations using NAO basis functions, an agreement  within 0.2 eV has been observed for the $G_0W_0$ benchmark calculation under "\textit{tier} 2" basis sets with respect to LAPW+lo result  for periodic systems.\cite{ren2021all} Meanwhile, the molecular BSE benchmark calculations showed that ``\textit{tier} 2" basis set is adequate for accurately capturing the electron-hole interaction of low-lying valence excited states when additional diffusive augmentation  Gaussian functions ``aug")\cite{papajak2011perspectives} are included in the basis set \cite{liu2020all}.  
These recent developments pave the way for the work presented here.
Building on our all-electron NAO-based $GW$ method for extended systems \cite{ren2021all} and all-electron NAO-based BSE method for isolated systems\cite{liu2020all,yao2022all}, 
we introduce a new all-electron NAO implementation of the BSE method for extended periodic systems in this work.

%\yk{Agreement of GW is wrt LAPW+lo?}

%\yk{There is a reference for aug basis sets, I beleive.}

%%%%%%%%%%%%%%%%%%%%%%%%%%%%%%
\section{3. All-electron Implementation with Numeric Atomic Orbitals Basis for Extended Periodic Systems}

Throughout this section, we utilize the following indices: $i, j, k, l$ for denoting occupied Kohn-Sham (KS) orbitals, and $a, b, c, d$ for denoting unoccupied KS orbitals. For the atomic orbital (AO) basis, we employ the indices $m$ and $n$, while the Greek letters $\mu, \nu, \alpha,$ and $\beta$ are used for auxiliary basis functions (ABFs) applied in the resolution of identity approach. $\mathbf{k}_1$ and $\mathbf{k}_2$ are used for k-point sampling in the BZ for KS orbitals while $\mathbf{q}$ is used for the grid sampling in the BZ for ABFs.

\subsubsection{3.1 Numeric Atomic Orbital Basis  Representation  }
NAO basis functions have the general form
%{xr{Now the abbreviation NAO has been defined previously} h
\begin{equation}
\varphi_n(\mathbf{r})=\frac{u_n(r)}{r}Y_{l,m_l}(\Omega)
\end{equation}
where $u_n(r)$ is the radial part and numerically  tabulated.
$Y_{l,m_l}(\Omega)$ are real-valued functions, comprised of either  the real parts $(m_l = 0,...,l)$ or the imaginary parts $(m_l=-l,...,-1)$ of the complex-valued spherical harmonics. 
The indices $l$ and $m_l$ are quantum numbers, describing the angular momentum quantities of spherical harmonic functions  $Y_{l,m_l}(\Omega)$ associated with the basis function index $n$. 

The definition of the NAO basis functions allows for using a wide range of shapes, including both analytically and numerically defined functions. This includes traditional quantum chemistry's analytically defined Gaussian-type or Slater-type orbitals. 
A key advantage of the NAO basis is the flexibility associated with choosing $u_n(r)$, and one can select numerical solutions for the  Schrödinger-like radial equations\cite{blum2009ab}
\begin{equation}\label{eq:nao_radial_equation}
    [-\frac{1}{2}\frac{d^2}{dr^2}+\frac{l(l+1)}{2r^2}+v_n(r)+v_{cut}(r)]u_n(r)=\epsilon_nu_n(r)
\end{equation}
%\xr{There should ba a factor of 1/2 in front of the term $l(l+1)/r^2$, which is also missing in the 2009 CPC paper.}

This Schrödinger-like equation includes a potential term, $v_n(r)$, which determines the primary behavior of $u_n(r)$, and another steeply increasing confining potential,  $v_{cut}(r)$. The confining potential can be chosen to ensure that each radial function, $u_n(r)$, decays smoothly and becomes strictly zero beyond a specific confinement radius. For a  comprehensive discussion on the NAO basis set, readers are referred to Ref. \cite{blum2009ab} .

For a periodic system, the Kohn-Sham (KS) orbital, denoted as $\psi_{i/a}^{\mathbf{k}}(\mathbf{r})$, can be represented as a linear combination of Bloch-adapted atomic orbitals as the basis set functions,
%\xr{it was said previously that the indices $n$, $m$ are reserved for NAO basis functions. In fact, they have been used for the excitation energy levels.} 
\begin{equation}\label{eq:mo_coeff}
           \psi_{i/a}^{\mathbf{k}}(\mathbf{r})=\sum_{m}\sum_{\mathbf{R}} e^{i\mathbf{k \cdot R}} c_{m,i/a}^{\mathbf{k}}\varphi_m(\mathbf{r}-\mathbf{\tau_m}-\mathbf{R}) 
           % \yk{not needed}\psi_{a}^{\mathbf{k}}(\mathbf{r})&=\sum_{m}\sum_{\mathbf{R}} e^{i\mathbf{k \cdot R}} c_{m,a}^{\mathbf{k}}\varphi_m(\mathbf{r}-\mathbf{\tau_m}-\mathbf{R}) 
\end{equation}
where $\varphi_m$ is the NAO basis function centered at the atomic position $\mathbf{\tau}_m$, from which the m-th atomic basis originates within the unit cell, and the sum runs over all unit cells $\mathbf{R}$ in the  Born–von Karman(BvK) super-cell. 

%\xr{BvK has not been defined before.}

\subsubsection{3.2  BSE in the Auxiliary Basis Set of Resolution of Identity}

Construction of the particle-hole kernel, through Eqs. \ref{eq:bse_kernel_V} and \ref{eq:BSE_kernel_W} is a major computational task. 
The direct evaluation of four-center integrals has historically posed challenges due to their significant computational and memory requirements in first-principles theory \cite{szabo2012modern}. The so-called Resolution of Identity (RI) approximation, also known as the density fitting, is a commonly employed method to alleviate the large computational cost in calculations with atom-centered orbitals like NAOs and Gaussians as basis functions.\cite{whitten1973coulombic,dunlap1979some}
Hartree-Fock \cite{vahtras1993integral,weigend2002fully} and other post-Hartree-Fock methods such as second-order Moller-Plesset perturbation theory (MP2) \cite{feyereisen1993use,weigend1998ri} and coupled cluster (CC) \cite{hattig2003geometry,schutz2003linear} often utilize the RI method. 
The RI approximation streamlines the calculation by reducing all four-center two-electron Coulomb integrals to pre-computed three- and two-center integrals.\cite{whitten1973coulombic,sierka2003fast}

\begin{comment}
For implementation of BSE for extended periodic systems, 
the choice of auxiliary basis typically depends on the pseudo-potential scheme, which may involve either the plane-wave \cite{albrecht1998ab,rohlfing1998electron,rohlfing2000electron,rocca2012solution,sander2015beyond} or Linear Augmented Plane Wave (LAPW) frameworks, also known as the mixed product basis \cite{sagmeister2009time} \yk{this sentence is not clear despite its potential importance. Does it belong here? RI is used in PW-based implementation?}.

\rz{This discussion is borrowed part from Xinguo's discussion in G0W0 paper. Plane wave or LAPW scheme can serve as aux basis for RI approximation? }
 
\end{comment}

The present all-electron NAO-based implementation also employs the RI approximation through constructing a set of NAO auxiliary basis functions to expand the products of two NAO orbitals, as described in Ref. \cite{ren2012resolution,ren2021all}.
For isolated systems (non-periodic case), the product of two NAO basis functions can be approximated within the RI approximation as a linear combination of auxiliary basis functions (ABF) as
\begin{equation}
\varphi_{m}^*(\mathbf{r})\varphi_{n}(\mathbf{r})=\sum_{\mu}C_{m,n}^{\mu}P_{\mu}(\mathbf{r-\tau_{\mu}})
\end{equation}
where $P_\mu(r)$ represents the $\mu$-th auxiliary basis function and $C_{m,n}^{\mu}$ is the expansion coefficient for the three-orbital (triple) expansion.
The expansion coefficient $C_{m,n}^{\mu}$ is given by
\begin{equation}\label{eq:RI-V}
C_{m,n}^{\mu}=\sum_{\nu}\langle m n|\hat{V}|\nu \rangle V_{\nu \mu}^{-1}
\end{equation}
where  $\langle m n|\hat{V}|\nu \rangle$  is the three center-Coulomb integral  given by 
\begin{equation}
\langle m n|\hat{V}|\nu \rangle=\iint\frac{\psi_{m}^*(\mathbf{r})\psi_{n}(\mathbf{r}) {P}_{\nu}(\mathbf{r}^{\prime})}{|\mathbf{r}-\mathbf{r}^{\prime}|}d \mathbf{r}d \mathbf{r}^{\prime}
\end{equation}
and $V_{\mu\nu}$ is the two-center Coulomb integral 
\begin{equation}
V_{\mu\nu}=\iint\frac{P_{\mu}(\mathbf{r}){P}_{\nu}(\mathbf{r}^{\prime})}{|\mathbf{r}-\mathbf{r}^{\prime}|}d \mathbf{r}d \mathbf{r}^{\prime}.
\end{equation}
The four-centered two-electron Coulomb integral, for instance, can be conveniently calculated as
\begin{equation}
\langle m n |\hat{V}|p q\rangle = \sum_{\mu \nu}C_{m,n}^{\mu} V_{\mu \nu} C_{p,q}^{\nu}
\end{equation}
This approach applicable in the non-periodic case is referred to as the ``RI-V" method in the subsequent discussion.

\subsubsection{3.3 Periodic systems}
For periodic systems, the products of Bloch-based atomic orbitals can be expanded using Bloch-based atom-centered Auxiliary Basis Functions (ABFs), 

\begin{equation}\label{eq:def_aux}
\varphi_{m}^{\mathbf{k+q}*}(\mathbf{r})\varphi_{n}^{\mathbf{k}}(\mathbf{r})=\sum_{\mu}^{N_{aux}}C_{m,n}^{\mu}(\mathbf{k+q},\mathbf{k})P_{\mu}^{\mathbf{q}*}(\mathbf{r}).
\end{equation}
where $N_{aux}$ represents the number of ABFs within each unit cell and the Bloch-based atom-centered ABFs,  $P_{\mu}^{\mathbf{q}}(\mathbf{r})$, are defined through Bloch theorem as \cite{bloch1929quantenmechanik}

\begin{equation}
P_{\mu}^{\mathbf{q}}(\mathbf{r})=\sum_{\mathbf{R}}P_{\mu}^{\mathbf{q}}(\mathbf{r-R-\tau_{\mu}})e^{i\mathbf{q}\cdot \mathbf{R}}
\end{equation} 

$C_{m,n}^{\mu}(\mathbf{k+q},\mathbf{k})$  is the atomic orbital (AO) based RI expansion coefficient, which depends on two independent Bloch wave-vectors, $\mathbf{k+q}$ and $\mathbf{k}$. 
Following  Ref. \cite{ren2012resolution} and \cite{ren2021all},
the matrix representation of the Coulomb operator $\hat{V}$ and static screened Coulomb operator in terms of the ABFs read 
\begin{equation}
\begin{split}
        V_{\mu\nu}(\mathbf{q})&=\iint\frac{P_{\mu}^{\mathbf{q*}}(\mathbf{r}){P}_{\nu}^{\mathbf{q}}(\mathbf{r}^{\prime})}{|\mathbf{r}-\mathbf{r}^{\prime}|}d \mathbf{r} d \mathbf{r}^{\prime}\\ W_{\mu\nu}\,(\mathbf{q})&=\iint P_{\mu}^{\mathbf{q}\ast}(\mathbf{r})W(\mathbf{r},\mathbf{r}^{\prime})P_{\nu}^{\mathbf{q}}(\mathbf{r}^{\prime})d \mathbf{r} d \mathbf{r}^{\prime}.
\end{split}
\end{equation}
With the definition of the screened Coulomb operator  $\hat{W}$, the matrix can be computed from the static dielectric matrix \cite{ren2021all} such that

\begin{equation}\label{eq:screen_aux}
        W_{\mu\nu}(\mathbf{q})=\sum_{\alpha \beta}V_{\mu \alpha}^{\frac{1}{2}}(\mathbf{q})\epsilon_{\alpha,\beta}^{-1}(\mathbf{q})V_{ \beta \nu}^{\frac{1}{2}}(\mathbf{q})
\end{equation}
where $V^{\frac{1}{2}}$ represents the square root of the Coulomb matrix $V$ and $\epsilon$ represents the symmetrized static dielectric function, whose matrix elements are computed as 
\begin{equation}\label{eq:dielectric_aux}
    \epsilon_{\mu\nu}(\mathbf{q})=\delta_{\mu\nu}-\sum_{\alpha\beta}V_{\mu \alpha}^{\frac{1}{2}}(\mathbf{q})\chi_{0,\alpha\beta}(\mathbf{q})V_{ \beta \nu}^{\frac{1}{2}}(\mathbf{q}).
\end{equation}
$\chi_0$ is the non-interacting static response function, according to the Adler-Wiser formula\cite{adler1962quantum,wiser1963dielectric},
\begin{equation}
\chi_0\left(\mathbf{r}, \mathbf{r}^{\prime}\right)=\sum_{i, a} \sum_{\mathbf{k}, \mathbf{q}}^{1 \mathrm{BZ}} 2 w_{\mathbf{k}} w_{\mathbf{q}} \frac{ \psi_{i}^{\mathbf{k}+\mathbf{q}*} (\mathbf{r}) \psi_{a}^{\mathbf{k}}(\mathbf{r}) \psi_{a}^{\mathbf{k} *}\left(\mathbf{r}^{\prime}\right) \psi_{i}^{\mathbf{k}+\mathbf{q}}\left(\mathbf{r}^{\prime}\right)}{\epsilon_{i}^{\mathbf{k}+\mathbf{q}}-\epsilon_{a}^{\mathbf{k}}}.
\end{equation}
%\xr{There are two terms in the frequency-dependent response function. Shouldn't there also be a second term here coming from swapping the indices?}
We need this response function given in the basis of ABFs, $\chi_{0,\alpha\beta}$, in Eq. \ref{eq:dielectric_aux}. 
To this end, we introduce the molecular orbital (MO) based RI expansion coefficients $\tilde{C}(\mathbf{k+q},\mathbf{k})$ such that

\begin{equation}\label{eq:def_aux_mo}
\psi_{i/a}^{\mathbf{k+q}*}(\mathbf{r})\psi_{j/b}^{\mathbf{k}}(\mathbf{r})=\sum_{\mu}^{N_{aux}}\tilde{C}_{i/a,j/b}^{\mu}(\mathbf{k+q},\mathbf{k})P_{\mu}^{\mathbf{q}*}(\mathbf{r}).
\end{equation}
where $\psi^{\mathbf{k}}$ are KS orbitals.
The MO-based expansion coefficients are related to the AO-based expansion coefficients by

\begin{equation}\label{eq:def_mo_coeff}
\begin{aligned}
      \tilde{C}_{i,j}^{\mu}(\mathbf{k_1},\mathbf{k_2})&=\sum_{m,n}c^*_{m,i}(\mathbf{k_1})c_{n,j}(\mathbf{k_2}) C_{m,n}^{\mu}(\mathbf{k_1},\mathbf{k_2})  \\
      \tilde{C}_{a,b}^{\mu}(\mathbf{k_1},\mathbf{k_2})&=\sum_{m,n}c^*_{m,a}(\mathbf{k_1})c_{n,b}(\mathbf{k_2}) C_{m,n}^{\mu}(\mathbf{k_1},\mathbf{k_2})  \\      
\end{aligned}
\end{equation}
where $c_{m,i/a}$ are molecular orbital (KS) coefficients which depends only on a single wave vector (see  Eq. \ref{eq:mo_coeff}) and $C_{m,n}^{\mu}(\mathbf{k_1},\mathbf{k_2})$ are the AO-based expansion coefficients (see Eq. \ref{eq:def_aux}).
Both the AO-based  and MO-based expansion coefficients depend on two momentum vectors $\mathbf{k_1}$ and $\mathbf{k_2}$.  
Then, the non-interacting response function $\chi_0$ in the auxiliary basis is given by
\begin{equation}
\chi_{0, \mu \nu}(\mathbf{q})=\sum_{ i,a} \sum_{\mathbf{k}} w_{\mathbf{k}} \frac{\tilde{C}_{i,a}^\mu(\mathbf{k}+\mathbf{q}, \mathbf{k}) \tilde{C}_{a,i}^\nu(\mathbf{k}, \mathbf{k}+\mathbf{q})}{\epsilon_{i}^{\mathbf{k}+\mathbf{q}}-\epsilon_{a}^{\mathbf{k}}}.
\end{equation}
Finally, the matrix elements of the Coulomb operator $\hat{V}$ and the static screened Coulomb operator $\hat{W}$ needed for constructing the BSE Hamiltonian (see Eq. \ref{eq:eq_A}) can be computed as
\begin{equation} \label{eq:eq_V2}
\begin{split}
\braket{i\mathbf{k_1} a\mathbf{k_1}|\hat{V}|j \mathbf{k_2} b\mathbf{k_2} }
= & \iint d\mathbf{r} d\mathbf{r'} \psi_{i,\mathbf{k_1}}(\mathbf{r}) \psi_{a,\mathbf{k_1}}^*(\mathbf{r})v(\mathbf{r},\mathbf{r'}) \psi_{j,\mathbf{k_2}}^*(\mathbf{r'}) \psi_{b,\mathbf{k_2}}(\mathbf{r'}) \\
= & \sum_{\mu\nu} \tilde{C}_{i,a}^{\mu*}(\mathbf{k_1},\mathbf{k_1}) V_{\mu\nu}(\mathbf{0}) \tilde{C}_{j,b}^{\nu}(\mathbf{k_2},\mathbf{k_2})
\end{split} 
\end{equation}
\begin{equation} \label{eq:eq_W2}
\begin{split}
\braket{ i\mathbf{k_1} j\mathbf{k_2}|\hat{W}|a\mathbf{k_1}b\mathbf{k_2}}
= & \iint d\mathbf{r} d\mathbf{r'} \psi_{i,\mathbf{k_1}}(\mathbf{r}) \psi_{j,\mathbf{k_2}}^*(\mathbf{r})W(\mathbf{r},\mathbf{r'}) \psi_{a,\mathbf{k_1}}^*(\mathbf{r'}) \psi_{b,\mathbf{k_2}}(\mathbf{r'}) \\
= & \sum_{\mu\nu} \tilde{C}_{i,j}^{\mu*}(\mathbf{k_1},\mathbf{k_2}) W_{\mu\nu}(\mathbf{k_2}-\mathbf{k_1}) \tilde{C}_{a,b}^{\nu}(\mathbf{k_1},\mathbf{k_2})
\end{split} 
\end{equation}

\subsubsection{3.3.a Local RI technique }
%Solving for the AO based  expansion coefficients $C_{m,n}^{\mu}(\mathbf{k_1},\mathbf{k_2})$ explicitly in the periodic system has posed a longstanding challenge. 
In contrast to the molecular case, where the RI-V method (Eq. \ref{eq:RI-V}) can be directly applied, solving for these coefficients in periodic systems presents significant difficulties. One major obstacle arises from the long-range nature of two-centered and three-centered integrals, necessitating the use of Ewald summation techniques in the integral construction. While notable progress has been made in recent years on addressing this challenge \cite{sun2017gaussian,ye2021fast}, it remains a highly nontrivial task to implement them. Additionally, it is worth noting that the computational cost of computing and storing AO triple coefficients scales as 
 $O(N_{aux}N_{b}^2N_{k}^2)$ where $N_b$ represents the number of basis functions, and $N_k$ represents the number of k-points, and thus the RI-V formalism is computationally quite expensive.
To address these issues, FHI-aims implementation utilizes the LRI (Local Resolution of Identity) approximation,, called RI-LVL in Ref.~\cite{ihrig2015accurate}.
Within the LRI approximation, the ABFs are used to expand the product of two NAOs are limited to those ABFs centered on the same two atoms on which the NAOs are centered.
In the quantum chemistry community, this two-center LRI scheme is also referred to as the Pair-Atom RI (PARI) approximation \cite{merlot2013attractive,wirz2017resolution}. 
In the context of periodic systems, the LRI approximation has been implemented for hybrid exchange-correlation  functionals for DFT\cite{levchenko2015hybrid,lin2020accuracy}, MP2\cite{ren2012resolution,zhang2019main}, RPA\cite{ren2012resolution,shi2024subquadratic} and GW \cite{ren2021all} methods within the NAO basis framework.  

In real space, the two NAOs, labeled as $m$ and $n$, can originate from different unit cells, denoted by two Bravais lattice vectors $\mathbf{R}_m$ and $\mathbf{R}_n$. The LRI approximation for periodic systems implies that:\cite{levchenko2015hybrid}
\begin{equation}\label{eq:RI-LVL}
\begin{aligned}
\varphi_m(\mathbf{r}-\mathbf{R}_m-\boldsymbol{\tau}_m) \varphi_n(\mathbf{r}-\mathbf{R}_n-\boldsymbol{\tau}_n) \approx &\sum_{\mu \in M} C_{m(\mathbf{R}_m), n(\mathbf{R}_n)}^{\mu(\mathbf{R}_m)} P_\mu(\mathbf{r}-\mathbf{R}_m-\boldsymbol{\tau}_m) + \\
&\sum_{\mu \in N} C_{m(\mathbf{R}_m), n(\mathbf{R}_n)}^{\mu(\mathbf{R}_n)} P_\mu(\mathbf{r}-\mathbf{R}_n-\boldsymbol{\tau}_n) 
\end{aligned}
\end{equation}
where $M$ and $N$ constitute the atoms on which AO basis functions $ \varphi_m$ and
$ \varphi_n$ are centered, and the summation over the ABFs is restricted to those ABFs centered on those atoms.
By minimizing the self Coulomb repulsion of the expansion error given by Eq. \ref{eq:RI-LVL}, the expansion coefficient can be determined as\cite{ihrig2015accurate}
\begin{equation}
C_{m(\mathbf{0}), n(\mathbf{R})}^{\mu(\mathbf{0})}=\begin{cases}
    \sum_{\nu \in\{M, N(\mathbf{R})\}}\langle m(\mathbf{0}), n(\mathbf{R})|\hat{V}|\nu\rangle \left(V^{M N}\right)_{\nu \mu}^{-1} & \text { for } \mu \in M \\
0 & \text { otherwise } \, .
\end{cases}
\end{equation} 
Instead of solving the inverse matrix of the entire two-electron Coulombic matrix as seen in Eq. \ref{eq:RI-V}, the LRI method computes the inverse of a local metric within the domain of $\nu \in{M, N(\mathbf{R})}$, centered either on the atom M in the original cell  $\mathbf{0}$ or on the atom N in the cell specified by $\mathbf{R}$ as shown in the above equation. 
Detailed discussion on the LRI approximation can be found in Ref. \cite{ihrig2015accurate}. 

To further derive the expansion coefficient in reciprocal space, we utilize the 
transnational symmetry property of periodic systems, i.e. $C_{m\left(\mathbf{R}_m\right), n(\mathbf{\mathbf{R}_n})}^{\mu(\mathbf{\mathbf{R}_n})}=C_{m\left(\mathbf{R}_m-\mathbf{R}_n\right), n(\mathbf{0})}^{\mu(\mathbf{0})} $, and Eq. \ref{eq:RI-LVL} can be transformed as
\begin{equation}
\begin{aligned}
       \varphi_m\left(\mathbf{r}-\mathbf{R}_m-\boldsymbol{\tau}_m\right) \varphi_n\left(\mathbf{r}-\mathbf{R}_n-\boldsymbol{\tau}_n\right) \approx &\sum_{\mu \in M} C_{m(\mathbf{0}), n\left(\mathbf{R}_n-\mathbf{R}_m\right)}^{\mu(\mathbf{0})} P_\mu\left(\mathbf{r}-\mathbf{R}_m-\boldsymbol{\tau}_m\right)+
    \\
    &\sum_{\mu \in N} C_{m\left(\mathbf{R}_m-\mathbf{R}_n\right), n(\mathbf{0})}^{\mu(\mathbf{0})} P_\mu\left(\mathbf{r}-\mathbf{R}_n-\boldsymbol{\tau}_n\right). 
\end{aligned}
\end{equation}
Through Fourier transformation, the product of Bloch-based atomic orbitals in LRI approximation can be derived as
\begin{equation}\label{eq:RI-LVL_derivation}\small
\begin{aligned}
\varphi_m^{\mathbf{k}+\mathbf{q} *}(\mathbf{r}) \varphi_n^{\mathbf{k}}(\mathbf{r}) & =\sum_{\mathbf{R}_m, \mathbf{R}_n} e^{-i(\mathbf{k}+\mathbf{q}) \cdot \mathbf{R}_m} e^{i \mathbf{k} \cdot \mathbf{R}_n} \varphi_m\left(\mathbf{r}-\mathbf{R}_m-\boldsymbol{\tau}_m\right) \varphi_n\left(\mathbf{r}-\mathbf{R}_n-\boldsymbol{\tau}_n\right) \\
&\approx \sum_{\mathbf{R}_m, \mathbf{R}_n} e^{-i(\mathbf{k}+\mathbf{q}) \cdot \mathbf{R}_m} e^{i \mathbf{k} \cdot \mathbf{R}_n}[\sum_{\mu \in M} C_{m(\mathbf{0}), n\left(\mathbf{R}_n-\mathbf{R}_m\right)}^{\mu(\mathbf{0})}P_\mu\left(\mathbf{r}-\mathbf{R}_m-\boldsymbol{\tau}_m\right)
\\
&+\sum_{\mu \in N} C_{m\left(\mathbf{R}_m-\mathbf{R}_n\right), n(\mathbf{0})}^{\mu(\mathbf{0})} P_\mu\left(\mathbf{r}-\mathbf{R}_n-\boldsymbol{\tau}_n\right)] \\
& =\sum_{\mu \in M}\left[\sum_{\mathbf{R}_m} e^{-i \mathbf{q} \cdot \mathbf{R}_m} P_\mu\left(\mathbf{r}-\mathbf{R}_m-\boldsymbol{\tau}_m\right) \sum_{\mathbf{R}_n} e^{i \mathbf{k} \cdot\left(\mathbf{R}_n-\mathbf{R}_m\right)} C_{m(\mathbf{0}), n\left(\mathbf{R}_n-\mathbf{R}_m\right)}^{\mu(\mathbf{0})}\right]
\\
& +\sum_{\mu \in N}\left[\sum_{\mathbf{R}_n} e^{-i \mathbf{q} \cdot \mathbf{R}_n} P_\mu\left(\mathbf{r}-\mathbf{R}_n-\boldsymbol{\tau}_n\right) \sum_{\mathbf{R}_m} e^{-i(\mathbf{k}+\mathbf{q}) \cdot\left(\mathbf{R}_m-\mathbf{R}_n\right)} C_{m\left(\mathbf{R}_m-\mathbf{R}_n\right), n(\mathbf{0})}^{\mu(\mathbf{0})}\right] \\
 &= \sum_{\mu \in M} C_{m(-\mathbf{k}-\mathbf{q}), n(\mathbf{0})}^{\mu(\mathbf{0})} P_\mu^{\mathbf{q *}}(\mathbf{r})+\sum_{\mu \in N} C_{m(\mathbf{0}), n(\mathbf{k})}^{\mu(\mathbf{0})} P_\mu^{\mathbf{q} *}(\mathbf{r}) .
\end{aligned}
\end{equation}
Here, as shown in the above equation, the terms $C_{m(-\mathbf{k}-\mathbf{q}),n(\mathbf{0})}^{\mu(\mathbf{0})}$ and $C_{m(\mathbf{0}), n(\mathbf{k})}^{\mu(\mathbf{0})}$ can be determined through the Fourier transformation of the real-space term $C_{m(\mathbf{R}),n(\mathbf{0})}^{\mu(\mathbf{0})}$ and $C_{m(\mathbf{0}),n(\mathbf{R})}^{\mu(\mathbf{0})}$. Therefore, by comparing Eq. \ref{eq:def_aux} and Eq. \ref{eq:RI-LVL_derivation}, we can obtain the atomic centered expansion coefficients in the reciprocal space using the LRI approximation as
\cite{ren2021all}

\begin{equation}
C_{m, n}^\mu(\mathbf{k}+\mathbf{q}, \mathbf{k})= \begin{cases} C_{m(-\mathbf{k}-\mathbf{q}),n(\mathbf{0})}^{\mu(\mathbf{0})} & \mu \in M \\ C_{m(\mathbf{0}), n(\mathbf{k})}^{\mu(\mathbf{0})} & \mu \in N \\ 0 & \text { otherwise }\end{cases}
\end{equation}

In summary, using the set of above working formula, one can efficiently calculate the AO-based expansion coefficients in the reciprocal space and subsequently, through a linear transformation, derive the MO-based expansion coefficients. 
This approach enables an efficient computation of the matrix elements for Coulombic and static screened Coulombic interactions in the BSE formalism, as expressed in Eqs. \ref{eq:eq_V2} and \ref{eq:eq_W2} within the LRI approximation. 
An important advantage of this approximation is that the AO-based expansion coefficients become dependent on only a single vector, either $-\mathbf{k}-\mathbf{q}$ or $\mathbf{k}$, within the BZ, rather than both simultaneously. As a result, the computational cost and memory storage requirements associated with the RI coefficients can be significantly reduced.

\subsubsection{3.3.b Singularity Treatment at $\Gamma$ point}
One outstanding technical challenge for periodic BSE calculation are the singularities that appear in the Coulomb term $V$ and the static screened Coulomb term $W$ of the BSE kernel at the $\Gamma$ point. In three-dimensional (3D) systems, the inherent nature of the bare Coulomb potential, characterized by the $1/r$ behavior, leads to a $1/q^2$ divergence as $q$ approaches 0 in the reciprocal space. 
Within the traditional plane-wave basis functions, $e^{i \textbf{G} \cdot \textbf{r}}$, the Coulomb operator is well-known to have a specific matrix form given by $V_{\textbf{G},\textbf{G}^\prime}(\textbf{q})=4\pi \frac{\delta (\textbf{G},\textbf{G}^\prime)}{|\textbf{q}+\textbf{G}|^2}$. This divergence is observed in the matrix element where both indices, $\textbf{G}$ and $\textbf{G}^\prime$, are equal to 0. This particular element is often referred to as the ``head" term of the matrix with indices $\textbf{G}$ and $\textbf{G}^\prime$. In physical terms, this term reflects the interaction energy of an infinitely extended, periodic array of charges of the same sign, which is infinite even per unit cell.
In the atom-centered ABF representation, this divergence carries over to the matrix elements between two nodeless s-type functions, resulting in a $1/q^2$ divergence. Similarly, between one nodeless s-type and one nodeless p-type function, it leads to a $1/q$ divergence. \cite{ren2021all}. 
The analytical form of the Coulomb operator can be expressed as

\begin{equation}
V_{\mu,\nu}(\textbf{q})=\frac{v_{\mu\nu}^{(2)}}{q^2}+\frac{v_{\mu\nu}^{(1)}}{q}+\bar{V}_{\mu,\nu}(\textbf{q})
\end{equation}
where $\bar{V}_{\mu,\nu}(\textbf{q})$ represents the analytic
part of the Coulomb operator as $q$ approaches 0 while $v_{\mu\nu}^{(2)}$ and $v_{\mu\nu}^{(1)}$ are the coefficients of the matrix elements exhibiting $1/q^2$ and $1/q$ asymptotic behaviors, commonly referred to as the ``head" and ``wing" terms, respectively. This divergence behavior of $V_{\mu\nu}(\textbf{q})$ also extends to the screened Coulomb matrix $W_{\mu\nu}(\textbf{q})$ in non-metallic systems. 

For addressing this issue, two numerical schemes are well known in the context of the Coulomb singularity within periodic HF calculation. 
The first scheme, known as the Gygi-Baldereschi (GB) scheme\cite{gygi1986self}, incorporates an analytically integrable compensating function to eliminate the diverging term and subtracts it separately. The second scheme, referred to as the Spencer-Alavi scheme \cite{spencer2008efficient}, employs a truncated Coulomb operator that avoids the Coulomb singularity, while ensuring systematic convergence to the correct limit as the number of k-point increases.  
In this study, we deal with the singularity issue associated with both the bare Coulomb and screened Coulomb operators by adopting a similar approach employed for $G_0W_0$ method by Ren, et al. \cite{ren2021all}.  A brief outline is provided here and, one can  find a more comprehensive procedure in Refs. \cite{levchenko2015hybrid,ren2021all}.
Our approach primarily consists of two main steps:

1. \textbf{Modified Spencer-Alavi Scheme:} To address the singularity problem of the bare Coulomb operator, a truncated Coulomb operator is introduced. In this way, the regularity of the bare Coulomb matrix as $\textbf{q}\to 0$ can be ensured. This truncated operator, denoted as $V^{cut}(\textbf{q})$, is obtained within the auxiliary basis by replacing the $\frac{1}{|r-r'|}$ term with a truncated form, represented as $v^{cut}(|r-r'|)$. The expression for $v^{cut}(|r-r'|)$ is given by 

\begin{equation}\label{v_cut}
v^{cut}(r) = \frac{{\text{erfc}(\gamma r)}}{r} + \frac{1}{2}\text{erfc}\left[\frac{{\ln(r)-\ln(R_{cut})}}{{\ln(R_{\omega})}}\right] \times \frac{{\text{erfc}(\gamma r)}}{r}
\end{equation}
where $R_{cut}$ represents the cutoff radius. This expression retains the short-range part of the Coulomb potential while rapidly suppressing the long-range part beyond $R_{cut}$. The value of $R_{cut}$ is determined as the radius of a sphere inscribed inside the Born–von Kármán (BvK) supercell. 
As the density of the k-point mesh increases, $R_{cut}$ gradually grows, allowing for the restoration of the full bare Coulomb operator. To optimize performance, the screening parameter $\gamma$ and the width parameter $R_{\omega}$ in Eq. \ref{v_cut} can be adjusted. In this work, we utilized the values that had been optimized and reported in previous work on hybrid XC functionals \cite{levchenko2015hybrid} and $G_0W_0$\cite{ren2021all} implementations.

2. \textbf{Proper Treatment of Symmetrized Dielectric Function}

In the calculation of the screened Coulomb interaction given by Eq.~\ref{eq:screen_aux}, the bare Coulomb interaction enters in the numerator through $V^{1/2}(\mathbf{q})$ and 
the denominator through the dielectric function $\epsilon(\mathbf{q})$. In the numerator, replacing the bare $V$ by
its truncated counterpart $V^{cut}$ works very well, but doing so for the dielectric function $\epsilon$ is not a good strategy as the screening property is not properly described. Therefore, following Ref.~\cite{ren2021all}, we adopted a mixed scheme where in the numerator the bare Coulomb interaction is truncated as is in the
case of exact-exchange calculations~\cite{levchenko2015hybrid}, where the full Coulomb operator is used to calculate the dielectric function. Note that the dielectric function is regular and finite everywhere in the BZ
except for $\mathbf{q}=0$ where the divergence of the bare Coulomb operator needs to be analytically treated. This
can be done most conveniently in the basis set representation of the eigenvectors of the Coulomb matrix, as discussed
previously in the literature \cite{friedrich2009efficient,friedrich2010efficient,jiang2013fhi}. Within such a representation, similar to the plane-wave case, the divergence in the bare Coulomb potential as $\mathbf{q} \rightarrow 0$ is cancelled by the corresponding asymptotic behavior of $\chi^0$, and the ``head" and ``wing" terms
of the dielectric function $\epsilon$ at $\mathbf{q}=0$ can thus be properly treated. 
Afterwards, one can transform $\epsilon(\mathbf{q}=0)$ back to the ABF representation. Via such a treatment,
the $\epsilon_{\mu\nu}(\mathbf{q})$ matrix becomes regular everywhere in the BZ, and can be numerically inverted.
The screened Coulomb matrix calculated via Eq.~\ref{eq:screen_aux} is thus also regular everywhere in the BZ and can
now be conveniently employed in setting up the BSE equation. 

We note that, in the present work, only the ``head" term of the dielectric function at $\mathbf{q}=0$ within the basis representation of 
the eigenvectors of the Coulomb matrix was explicitly treated, while the ``wing" term, which has a secondary effect, was left untouched. 
This is because
a proper treatment of the ``wing" term was not available when the present research work was conducted. Very recently, a rigorous treatment of the ``wing'' term became available in FHI-aims code as part of a separate, ongoing study, and preliminary test calculations show that this entails a blue shift of the $G_0W_0$ band gap by 0.03 eV for Si and 0.13 eV for MgO. Therefore, we do not expect a significant influence of the ``wing" correction on the BSE@$GW$ results presented in the present work. Further investigation of this issue
will be presented in future work.

\section{4. Demonstration of NAO Implementation and Convergence}

%\vb{Would it make sense to place the Si and MgO results in the same section? This would perhaps allow us to avoid some otherwise necessary repetition?}

The new all-electron NAO-based periodic BSE method is implemented within the FHI-aims code\cite{blum2009ab,blum2022fhi}.  
For the BZ sampling, we employ an even-sampled $\Gamma$-centered grid with equally spaced $n_1 \times n_2 \times n_3$ k-points. The integration grid for the NAO basis for single-particle matrix elements and other real-space integrals employs FHI-aims' "tight" settings. The detailed choice and convergence  of the NAO basis set and auxiliary basis set are discussed in subsequent subsections. 
We use crystalline silicon (Si), 2-atom primitive cell, as an example to demonstrate our implementation, particularly focused on the NAO basis set, auxiliary basis set, and the BZ sampling. We also provide a direct comparison of our all-electron NAO results to  those obtained using the traditional PlaneWave Pseudopotential (PW-PP) approach implemented in the BerkeleyGW \cite{deslippe2012berkeleygw} code and Quantum Espresso \cite{giannozzi2020quantum} code.

%\vb{Mention MgO?}

The computational procedure starts with Kohn-Sham (KS) DFT calculations using the local density approximation (LDA) in the Perdew-Wang (PW) parameterization \cite{perdew1992accurate}. Subsequently, $G_0W_0$ calculations are performed on top of the KS orbitals and energies to obtain the quasi-particle energies. The BSE calculation is then performed with the results of the $G_0W_0$ calculation.  
For the frequency-dependent dielectric function, we utilize 80 frequency points in the Pade approximation for analytic continuation.  In constructing the BSE Hamiltonian, we used 4 valence bands and 6 conduction bands. (this choice is for Si and it would need separate validation for other systems).
%\vb{Is a validation possible? Would the high lying energy range be affected?} 

%\rz{Put in Introduction?}

%\rz{How well response converges with NAO? Energy windows/band gap(published in Xinguo's paper) Much better than molecule!}

%\rz{How well converges with BSE?}

%\rz{Molecular BSE? Energy window converges well.}

%%%%%%%%%%%%%%%%%%%%%%%%%%%%%%%%
\subsection{4.1 Convergence of Auxiliary Basis Set}

As discussed in Section 3.3, our periodic BSE method is implemented based on the LRI approximation, thus the accuracy depends on the quality of the auxiliary basis functions (ABFs). 
In the FHI-aims code, standard ABFs are generated from the one-electron orbital basis set (OBS) employed in the preceding KS calculations, summarized in Figure 1 of Reference \cite{ihrig2015accurate}. 
To construct the auxiliary basis from OBS, the radial components are derived directly from the products of the one-electron orbital basis sets, and then Gram-Schmidt orthogonalization is applied to eliminate linear dependencies within the auxiliary basis \cite{ren2012resolution}. 
In the LRI approximation, a larger number of ABFs is required than in the non-local RI-V scheme, since high angular momentum components of the pair density to be expanded must be accounted for by ABFs on the same two centers on which the orbital basis functions in questions are centered. \cite{ihrig2015accurate}. One practical strategy to create accurate auxiliary basis sets for LRI is to supplement the OBS with additional functions of higher angular momenta. It is important to note that these additional orbital basis functions are exclusively employed in the construction of ABFs and do not participate in the preceding self-consistent KS-DFT calculations. In the input files of FHI-aims, the supplemental functions to the OBS that generate the extended ABF basis set are denoted by the keyword \texttt{for$\_$aux}.

%\yk{In the FHI-aims input file, these supplementary functions are designated with a $for\_aux$ tag, indicating their sole purpose in generating auxiliary functions.}
Following the nomenclature introduced in Ref. \cite{ihrig2015accurate}, we refer to the set of OBS plus the additional supplemental functions as the enhanced orbital basis set (OBS+). It was previously demonstrated that for an OBS containing at least up to f functions, the inclusion of an extra 5g hydrogenic function in the enhanced orbital basis set (OBS+) can yield an ABF basis set with sufficient accuracy for various computational methods, including Hartree-Fock (HF), MP2, and RPA, in molecular calculations  \cite{ihrig2015accurate} Furthermore, based on  benchmark calculations of periodic $G_0W_0$, it was found that the inclusion of $4f$ or even $5g$ hydrogenic functions in the OBS+ is necessary to achieve convergence of the auxiliary basis.\cite{ren2021all} 
To assess the convergence of the auxiliary basis for the periodic BSE with the LRI scheme, we compute the absorption spectrum for crystalline silicon (Si) using an $8\times 8 \times 8$ BZ sampling. 
In defining augmented hydrogenic functions within the enhanced orbital basis set (OBS+),  an additional parameter, $Z$ is introduced, which represents an 
an effective charge for a hydrogen-like generating potential $v_n(r)$ in Eqn. \ref{eq:nao_radial_equation} and governs the shape and spatial extent of  the solution to the radial Schr{\"o}dinger equation \cite{blum2009ab,ihrig2015accurate}. In this work, we set $Z$=0 as the default value, resulting in the utilization of spherical Bessel functions, confined by the confining potential introduced in Eqn.  \ref{eq:nao_radial_equation}, for constructing the auxiliary basis.

\begin{figure}
    \centering
    \includegraphics[width=0.6\textwidth]{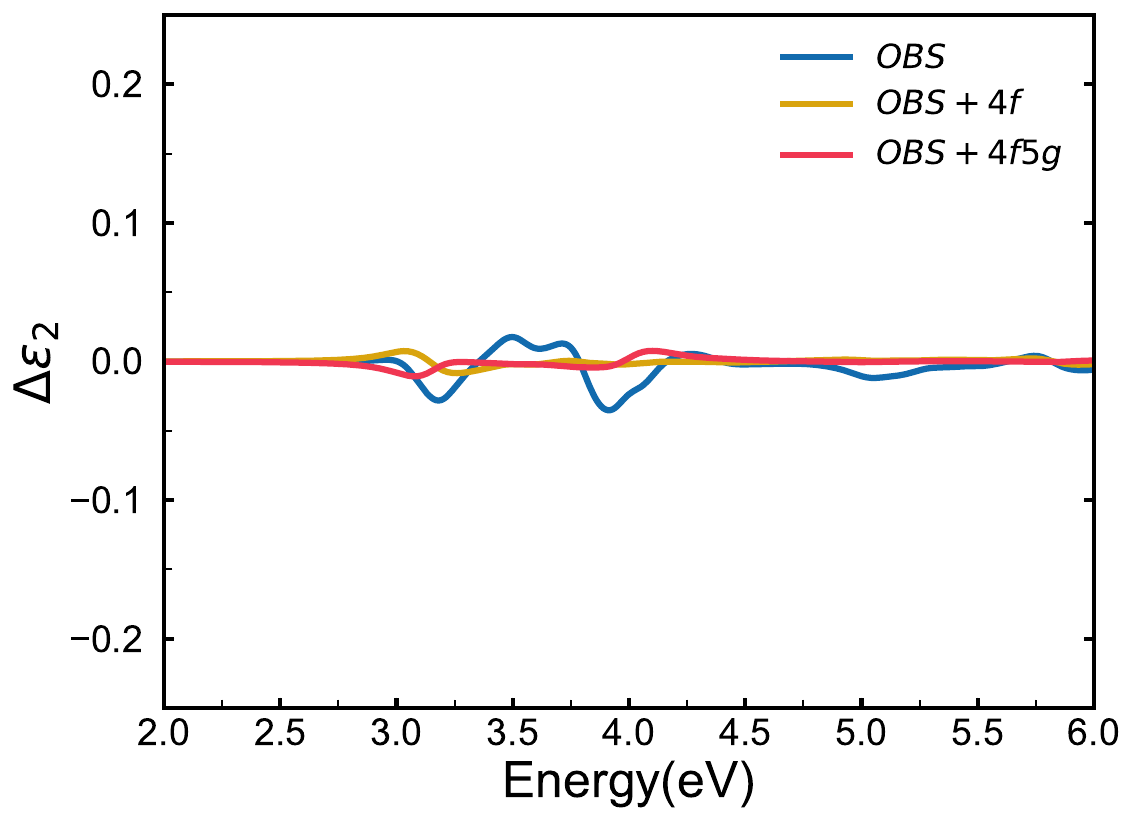}
    \caption{Relative error $\Delta \epsilon_2$ of the absorption spectrum of silicon (Si) from $BSE@G_0W_0$ calculations using different auxiliary numerical atomic orbitals (NAO) basis. The standard reference result is performed with the OBS+4f5g6h auxiliary NAO basis. All calculations are performed with $8\times 8\times 8$ $\Gamma$-centered BZ sampling and  \textit{tier} 2 NAO basis set.\cite{knuth2015all}}
    \label{fig:Si_Aux_error}
\end{figure}

As depicted in Figure \ref{fig:Si_Aux_error}, we employ the exhaustive OBS+4f5g6h result as the reference standard and compute the relative errors for OBS, OBS+4f, and OBS+4f5g. \cite{knuth2015all}
In terms of the optical energy gap, given by the lowest BSE eigenvalue, using the regular OBS to construct ABFs yields the value of 3.083 eV, which is already in excellent agreement with the reference value, differing only by 1 meV. 
However, for the features spanning from 3 eV to 8 eV in the optical absorption spectrum, the ABFs derived from the \textit{tier} 2 OBS only lead to exhibits a relatively large error in the absorption peak intensity as seen in Figure \ref{fig:Si_Aux_error}.  Introducing additional OBS+ auxiliary basis functions as in OBS+4f and OBS+4f5g make the LRI-related error negligible as they are converged with respect to the reference OBS+4f5g6h result (see Figure \ref{fig:Si_Aux_error}).
We note that, when comparing the BSE convergence with the $G_0W_0$ convergence test presented in Ref. \cite{ren2021all}, the sensitivity to the auxiliary basis in the BSE calculations is less pronounced than in the $G_0W_0$ calculations.  The primary difference arises from how the screened interaction is handled differently in BSE and $G_0W_0$ calculations. In BSE calculations, the screened interaction among electron-hole pairs near the Fermi level is important. In contrast, the self-energy evaluation in $G_0W_0$ calculations involve the screened interaction for higher-energy orbitals, necessitating a more extensive set of delocalized basis functions.

To summarize, we here demonstrated the robustness of the LRI scheme within our periodic BSE framework. To achieve the full convergence of auxiliary basis set, OBS+4f or more diffusive spherical Bessel functions with a confinement potential are found necessary. 
Thus, in all simulations presented in this work, OBS+4f is employed as the default setting for the auxiliary basis.

%%%%%%%%%%%%%%%%%%%%%%%%%%%%%%%%
\subsection{4.2 Convergence of NAO Basis set }

In terms of NAO basis sets convergence, very high precision can be reached in all-electron ground-state Density Functional Theory (DFT) calculations when only occupied KS states need to be evaluated \cite{blum2009ab,jensen2017elephant}. However, similar to the case of Gaussian Type Orbitals (GTOs) and other atom-centered basis sets, larger basis sets may be needed when applied to correlated calculations such as MP2 and RPA. The standard FHI-aims-2009 NAO basis set series ("\textit{tier} n" basis), while originally designed for ground-state DFT calculations, can actually produce acceptable results also for molecular MP2 and RPA calculation when counterpoise corrections are employed. \cite{ren2012resolution}. Enhanced accuracy can be achieved by employing the alternative, so-called valence-correlation consistent (VCC) NAO-VCC-$n$Z basis sets \cite{zhang2013numeric}, which allow for results to be extrapolated to the complete basis set (CBS) limit through a two-point extrapolation procedure \cite{zhang2019main}. 
In densely packed solids, however, it was observed that the original NAO-VCC-$n$Z basis sets, initially designed for molecules, lead to overlap matrices with impractically large condition numbers compared to the \textit{tier} n basis sets, and numerical instabilities in standard linear algebra (specifically, eigenvalue solutions) can be a consequence.
To address this limitation, Zhang et al. \cite{zhang2019main} optimized these basis sets by eliminating the so-called 'enhanced minimal basis' and tightening the cutoff radius of the basis functions. The resulting basis sets, referred to as localized NAO-VCC-$n$Z (loc-NAO-VCC-nZ) here, have been demonstrated to yield accurate MP2 and RPA energies for simple solids when used in conjunction with an appropriate extrapolation procedure in Ref. \cite{zhang2019main} Separately, in BSE@$G_0W_0$ benchmarks for molecules, our team observed excellent numerical convergence with the \textit{tier} n basis sets, augmented with two extended Gaussian orbital basis functions from Dunning's augmented correlation-consistent basis sets.\cite{liu2020all} 
To assess basis set convergence in the present work, we conducted calculations of the silicon (Si) absorption spectrum using the BSE@$G_0W_0$ method with two different sets of basis functions: (a)  \textit{tier} n ($n=1,2,3$) and (b) loc-NAO-VCC-$n$Z ($n=2,3,4$). These basis set convergence tests were performed using a $7\times 7\times 7$ $\Gamma$-centered BZ sampling.

\begin{figure}
    \begin{subfigure}[t]{0.50\textwidth}
            %\centering
			\includegraphics[width=\textwidth]{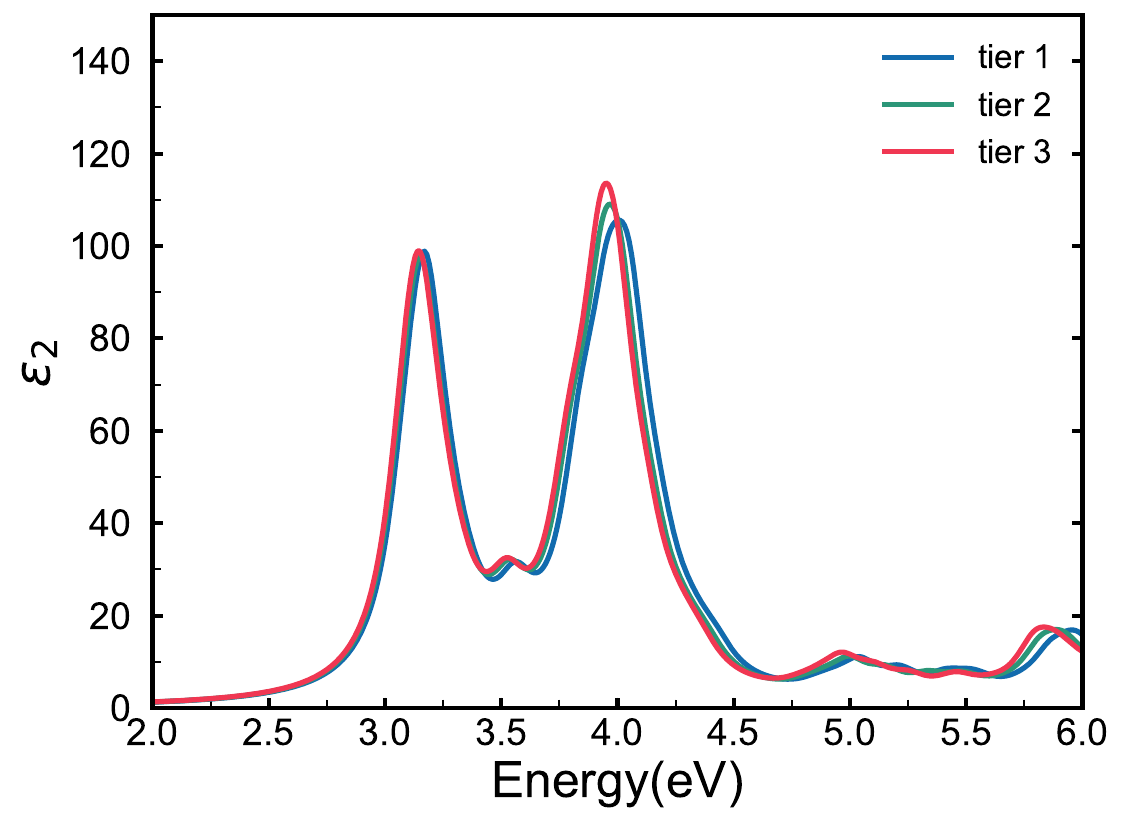} 
            \caption{}        
    \end{subfigure}
    \\
    \begin{subfigure}[t]{0.50\textwidth}
            %\centering
			\includegraphics[width=\textwidth]{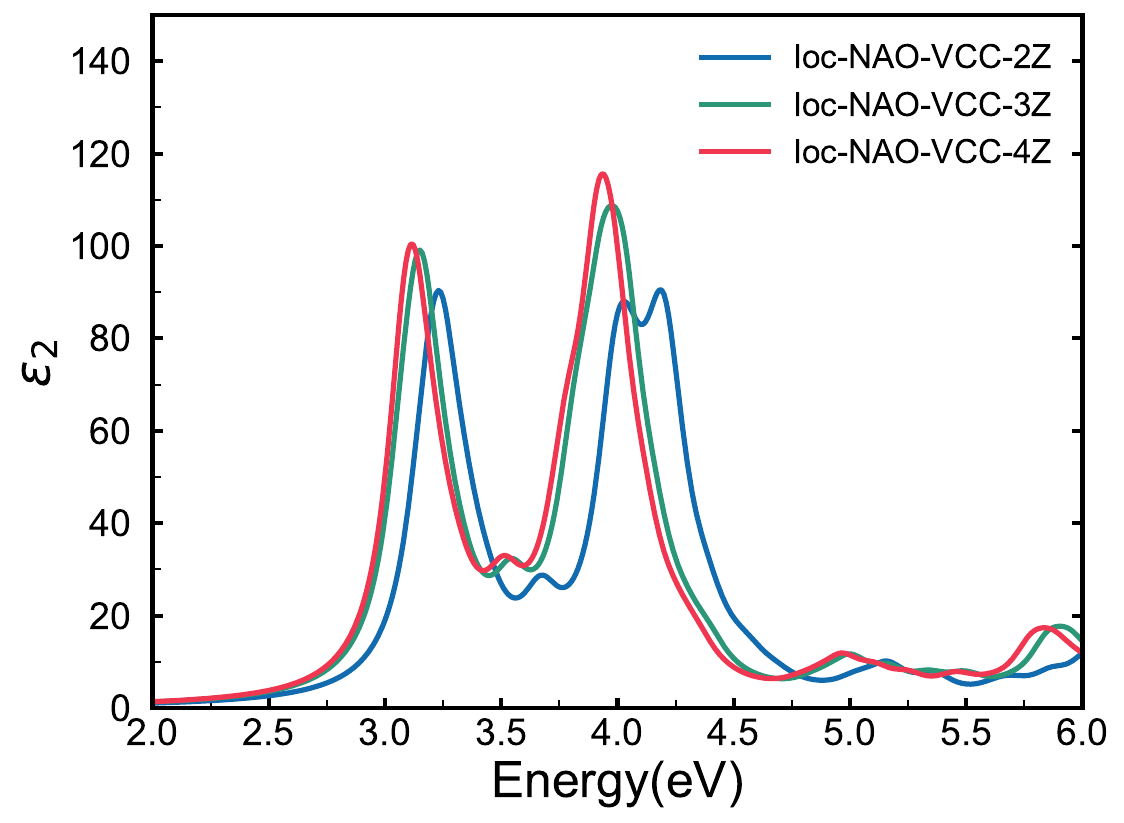} 
            \caption{}        
    \end{subfigure}
    \caption{Convergence of the absorption spectrum of Si from $BSE@G_0W_0$ calculation with respect to different NAO basis sets. A Lorentzian broadening of $\eta$=0.10~eV is used. The calculations are performed with $7\times 7\times 7$ $\Gamma$-centered BZ sampling and  (a) \textit{tier}  n $(n=1,2,3)$ and (b) loc-NAO-VCC-nZ $(n=2,3,4)$  NAO basis sets complemented by the OBS+4f auxiliary NAO basis set.
    }
    \label{fig:Si_basis}
\end{figure}

As shown in Figure \ref{fig:Si_basis} (a), the optical spectrum converges quickly in the \textit{tier} n basis set, with only a marginal shift in excitation peaks as we increase the basis set size from \textit{tier} 1 to \textit{tier} 3. For instance, the energy of the first peak changes by only 0.021eV from 3.173 eV to 3.152 eV as we move up from \textit{tier} 1 to \textit{tier} 3. Conversely, when employing the loc-NAO-VCC-$n$Z basis set, the absorption spectrum for the 2Z basis set is not fully converged as seen in Figure \ref{fig:Si_basis} (b). 
There is some qualitatively incorrect behavior as evidenced, for example, in the erroneous prediction of multiple peaks around 4 eV as well as the sizable blue-shift of the first excitation at 3.238 eV. 
These erroneous features can be attributed to the omission of the ``enhanced minimal basis" in the loc-NAO-VCC-2Z basis, rendering it insufficient for accurately describing the occupied states in the preceding self-consistent field (SCF) calculations. 
At the same time, the larger basis sets in this series, loc-NAO-VCC-3Z and loc-NAO-VCC-4Z basis sets, yield converged results, and they also agree well with the converged results using the ``\textit{tier} n" default basis sets of the FHI-aims code. 
To summarize, with the largest NAO-VCC-4Z and \textit{tier} 3 basis sets, we find precise agreement within 30 meV for the BSE@$G_0W_0$ absorption spectrum. Furthermore, even the smaller \textit{tier} 1 and \textit{tier} 2 NAO basis sets lead to results that are remarkably well converged. This behavior is a significant and highly promising success. In particular, this behavior is qualitatively different from the small-molecule case \cite{liu2020all}, in which extended augmentation functions were shown to be needed, as is standard in molecular quantum chemistry. We attribute this encouraging difference to the molecular case to the fact that in solids, there is a significantly higher density of basis functions that are non-zero at any given point, compared to finite molecules. The reason is the overlap of basis functions associated with neighboring unit cells or, equivalently, the basis functions with different Bloch phase factors in the BvK cell. In our view, this higher density of non-zero basis functions lends itself to a more finely resolved description of the two-particle correlation function than is possible in molecules, where describing the two-particle response is problematic especially in regions that are relatively distant from the atoms (covered by augmentation functions). In summary, it seems that FHI-aims' standard NAO basis sets can be used for well converged BSE@$GW$ calculations in solids. In the subsequent sections, we employ the default \textit{tier} 2 basis set of the FHI-aims code as specified  in Reference \cite{ren2021all}.

%%%%%%%%%%%%%%%%%%%%%%%%%%%%%%%%%
\subsection{4.3 Convergence of Brillouin Zone (BZ) Sampling }

One of the most important and also challenging aspects of calculating the optical absorption spectrum of extended condensed matter systems is
achieving the convergence with respect to the BZ sampling\cite{deslippe2012berkeleygw,kammerlander2012speeding,sander2015beyond,gillet2016efficient}. 
This is a particularly important consideration for many inorganic solids in which there exist strong band dispersion and the band gap is indirect. 
To accelerate the convergence of the absorption spectrum in BSE calculations, the BZ sampling with a randomly shifted k-grid around the $\Gamma$ point has been used for practical calculations as discussed in the literature \cite{rohlfing2000electron,schmidt2003efficient, marini2009yambo, rocca2012solution}. 
For the purpose of benchmarking our new NAO-based BSE implementation, we do not consider such accelerated k-point sampling techniques in this work, but we focus on the convergence behavior of the direct equally-spaced BZ sampling centered at $\Gamma$ point. 
To generate quasi-particle (QP) energies for BSE calculation, we employ $G_0W_0$ calculation for all k-grids instead of applying a consistent scissors shift to Kohn-Sham orbital energies as is sometimes done in the literature.
In the previous benchmark on the NAO-based $G_0W_0$ method\cite{ren2021all}, it was found that $7\times 7\times 7$ BZ sampling for the dielectric matrix is adequate to achieve convergence of QP energies within 5 meV. 

\begin{figure}
    \centering
    \includegraphics[width=0.6\textwidth]{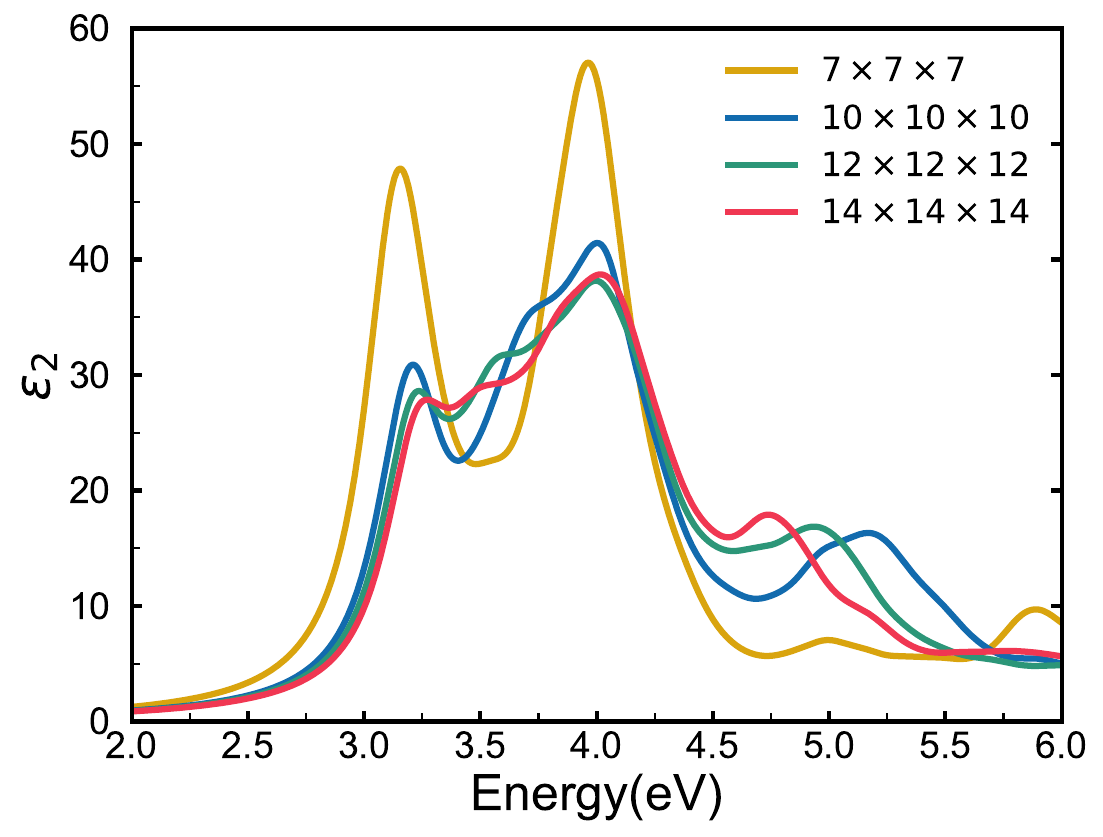}
    \label{fig:Si-BZ}
    \caption{Absorption spectrum of Si with  $\Gamma$-centered BZ sampling of dimensions $n \times n \times n$ ($n=7, 10, 12, 14$) calculated using the \textit{tier} 2 NAO basis set and OBS+4f auxiliary NAO basis set. A Lorentzian broadening of $\eta= 0.15 eV$ is used.  
     } 
    %\yk{You need to say BerkeleyGW, explicitly.} 
    %The $G_0W_0$ benchmark calculations employed two different numerical strategies for the frequency integration of the self-energy: the Generalized Plasmon-Pole (GPP) model \cite{hybertsen1986electron} and the Contour Deformation method for full-frequency calculations (full-freq).
    \label{fig:Si_BZ}
\end{figure}

Figure \ref{fig:Si_BZ} shows the optical absorption spectrum of crystalline silicon, utilizing $\Gamma$-centered uniform BZ sampling with $n \times n \times  n$ where $n=7, 10, 12, 14$. 
Neither the shape of the absorption spectrum nor the excitation energies are converged when employing the $7\times7\times7$ BZ sampling, despite the convergence observed for the $G_0W_0$ QP energies\cite{ren2021all}. 
With increased BZ samplings, the absorption spectrum generally becomes blue-shifted, particularly for the first absorption peak. 
At the same time, the peak at around 5.0 eV is seen to red-shift, and the spectrum does not converge uniformly at all energies. 
In general, the shape of the absorption spectrum tends to converge with the increased BZ sampling from $n=10$ to $n=14$. At the same time, we note that achieving the complete convergence of the absorption spectrum for cystalline silicon, as discussed in Ref. \cite{sander2015beyond}, may necessitate an exceedingly fine sampling of the BZ to the value of $n=40$. 

Our current implementation does not exploit space group symmetry, k-space interpolation strategies, or similar simplifications. Therefore, our current computational resources do not allow us to achieve the full convergence with with a significantly denser BZ sampling at this time. We intend to explore explore incorporating enhanced BZ sampling techniques in our future research effort, such as coarse-grained k-grid interpolation \cite{deslippe2012berkeleygw}, Wannier interpolation \cite{kammerlander2012speeding}, and other related schemes \cite{sander2015beyond,gillet2016efficient}.

%%%%%%%%%%%%%%%%%%%%%%%%%%
\subsection{4.4 Comparison with Planewave Basis-Set Result}

Having examined the convergence behavior of the BSE calculation for our all-electron NAO based implementation, we now turn to the comparison to
the BSE calculation based on the traditional plane-wave pseudopotential (PW-PP) implementation. 
We performed the PW-PP based BSE@$GW$ calculation using the well-documented BerkeleyGW package \cite{deslippe2012berkeleygw}. The BSE@$GW$ calculation was performed on top of the DFT calculation using the Quantum Espresso \cite{giannozzi2009quantum,giannozzi2017advanced} code. We used the local density approximation (LDA) to the exchange-correlational functional in the DFT calculation as the starting point\cite{perdew1992accurate}.
Further details of computational settings are provided in the Supporting Information. 

\begin{figure}
    \centering
    \begin{subfigure}{0.45\textwidth}
        \centering
        \includegraphics[width=\textwidth]{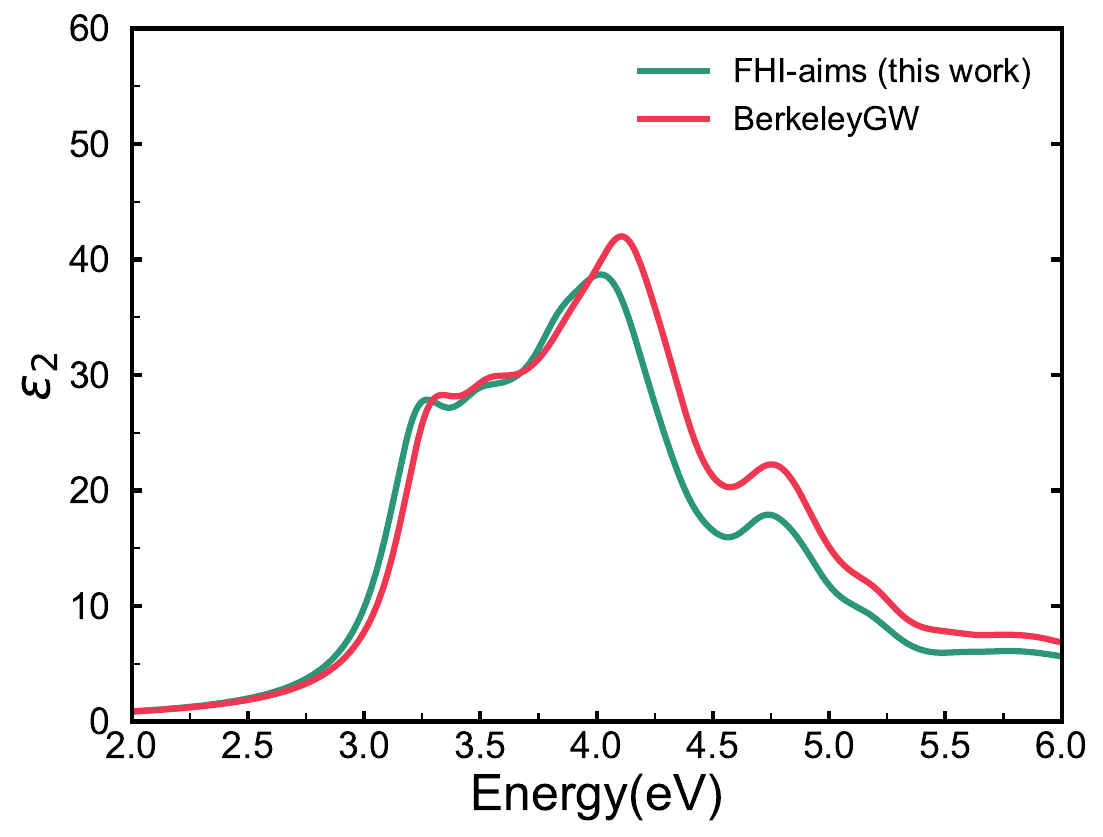}
        \label{fig:Si-abs-comp}
        \caption{}
    \end{subfigure}
    \hfill
    \begin{subfigure}{0.45\textwidth}
        \centering
        \includegraphics[width=\textwidth]{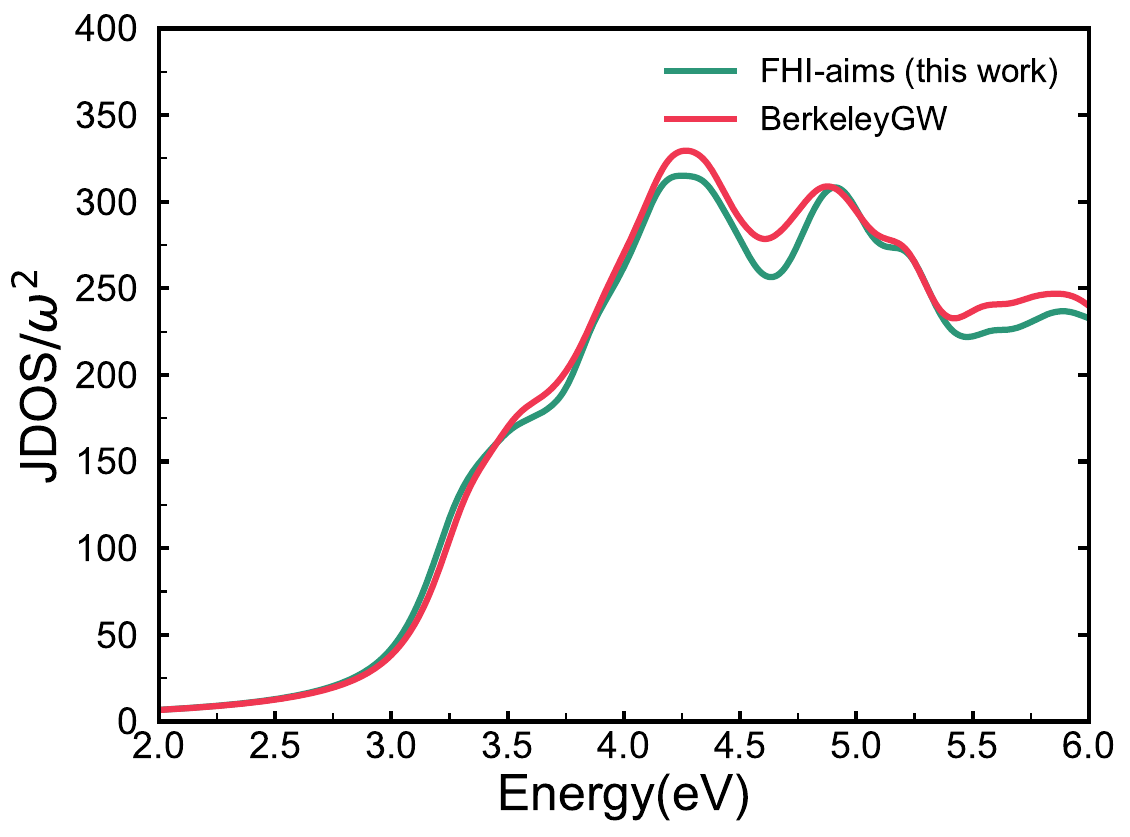}
        \label{fig:Si-JOS-comp}
        \caption{}
    \end{subfigure}
    \caption{Comparison of the (a) optical absorption spectra and (b) joint density of state (JDOS) divided by $\omega^2$ for Si between our NAO implementation in FHI-aims code and the PW-PP implementation in BerkeleyGW code, performed with a $14 \times 14 \times 14$ $\Gamma$-centered BZ sampling.  The FHI-aims calculation uses the \textit{tier} 2 NAO basis set and OBS+4f auxiliary NAO basis set. A Lorentzian broadening of $\eta= 0.15 eV$ is used.}
    \label{fig:Si-comp}
\end{figure}

In Figure \ref{fig:Si-comp}(a), we present the optical absorption spectrum of crystalline silicon, using a $14\times 14\times 14$ $\Gamma$-centered BZ sampling. Both spectra use the same Lorentzian broadening of 0.15 eV. 
Note that Contour Deformation (not the widely-used generalized plasmon-pole  model \cite{hybertsen1986electron}) technique is employed here for the numerical (frequency) integration of the self-energy in the $G_0W_0$ calculation with BerkeleyGW, in order to compare the two implementations on a similar footing. 
Our NAO-based BSE$@G_0W_0$ results closely match the PW-PP result in terms of the peak positions while some differences in the amplitude are observed at around 4.0-5.0 eV.  
We note that HOMO-LUMO QP energy gap from the $G_0W_0$ calculations agree very closely between our NAO-based result and the PW-PP result within 6 meV. 
BSE eigenvalues (i.e. excitation energies) from our NAO-based calculation also closely match with the PW-PP results, exhibiting an average difference of 28 meV. 

To better understand the origin of the observed differences in the optical absorption spectrum, Figure \ref{fig:Si-comp}(b) 
shows the comparison using the scaled joint density of states (JDOS),
\begin{equation}
    JDOS(\omega)/\omega^2=\frac{16\pi^2e^2}{\omega^2}\sum_{m} \delta(\omega-\omega_m)
\end{equation}
where $\omega_m$ is the energy of excited state $m$ from BSE calculation.
Instead of having the transition amplitudes convoluted as part of the calculation of the absorption spectrum, the JDOS allows for a direct comparison of electronic transitions as a function of the excitation energy. 
As seen in Figure \ref{fig:Si-comp}(b), our NAO-based BSE excitation energies closely align with the PW-PP results, except in the energy range of 4.0-4.5 eV, where a noticeble deviation is observed. 
This rather small deviation in this energy range has a significant impact on the differences observed in the absorption spectrum as seen in Figure \ref{fig:Si-comp}(a). 
Given the fully converged basis set for this particular case of crystalline silicon as discussed above, these differences are attributed to the difference between the two implementations, e.g. due to the difference in the frequency integration treatments utilized in the $G_0W_0$ calculation. In our benchmark calculation, the contour deformation  technique is employed for the numerical (frequency) integration of the self-energy in the $G_0W_0$ calculation using BerkeleyGW code, whereas Pade approximation is utilized in FHI-aims for analytic continuation\cite{vidberg1977solving} .

%\vb{Can this be explained? ... in principle, a complete comparison should also involve a comparison of the GW band structures produced by both codes. Can we say something about the expected agreement?}

%It's important to note that we employed two distinct strategies for the numerical (frequency) integration of the self-energy in the $G_0W_0$ benchmark calculation in BerkeleyGW, namely, the Generalized Plasma-Pole (GPP) model \cite{hybertsen1986electron} and Contour Deformation (CD).

%It's evident that the BSE absorption spectrum exhibits a dependence on the quasiparticle (QP) energies for different approaches. In comparison to the GPP model, the CD result shows a red-shift in the absorption spectrum. Moreover, the CD result closely aligns with our NAO result in terms of peak position, albeit some differences in amplitude persist.

%Similarly, the GPP result also exhibits good agreement, differing by approximately 98 meV in HOMO-LUMO quasi-particle energies. This is mainly attributed to the semiconductor nature of silicon, where electrons are relatively delocalized in the lowest and highest conduction bands, which is a typical scenario for obtaining accurate results with the GPP model. \cite{larson2013role}

%Consequently, BSE eigenvalues based on Contour Deformation align more closely with the AIMS result, with an average difference of 26 meV, which is smaller than the GPP result (82 meV). In conclusion, it can be inferred that quasi-particle energies contribute significantly to the observed differences in periodic BSE calculations, while the remaining disparities arise from variations in the treatment of core electrons in both all-electron and pseudo-potential calculations. 

\section{5. Application to Excitons with Large Binding Energy: MgO}
The BSE formalism explicitly solves for the two-particle correlation function, making it an ideal approach for studying extended systems with strongly excitonic character. 
In this section, we compare our all-electron NAO-based implementation of BSE@$GW$ with other BSE@$GW$ calculations reported in literature for crystalline magnesium oxide (MgO). 
%First, we consider the crystalline MgO because of its large exciton binding energy with this prototypical oxide material. Second, the dependence of the exciton binding energy on the atomistic disorder of water molecules in ice (Ic) is used to asses the performance of our all-electron implementation.

MgO is one of the extensively investigated oxides\cite{maruyama2009aa, yang2011magnetism}. 
Developing a comprehensive understanding of its spectroscopic features from first principles necessitates an accurate modeling of electronic excitation of this extended material with a large exciton binding energy.\cite{begum2021theoretical}
MgO also serves as an ideal benchmark system for studying optical properties using many-body perturbation theory, and  excellent agreement with experimental measurement has been reported \cite{begum2021theoretical, wang2004electronic, schleife2006first}.
We compare our all-electron NAO implementation of BSE@$GW$ with two other corresponding calculations based on the planewaves with the projector-augmented-waves method (PW+PAW) and linearized augmented planewave + local orbital (LAPW+lo) formalism reported in Ref. \cite{begum2021theoretical}.

\begin{figure}
    \centering
    \begin{subfigure}{0.45\textwidth}
        \includegraphics[width=\textwidth]{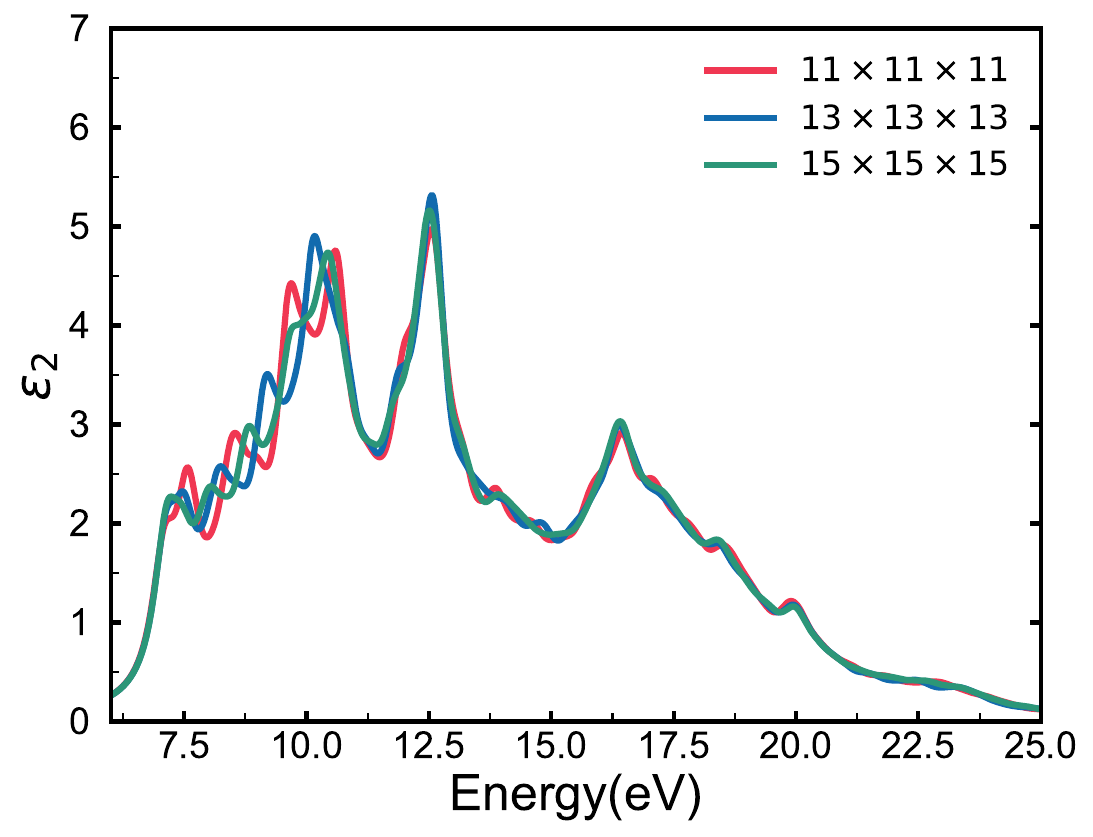}
        \caption{}
        %\label{fig:MgO-BZ}
    \end{subfigure}
    \hfill
    \begin{subfigure}{0.45\textwidth}
        \includegraphics[width=\textwidth]{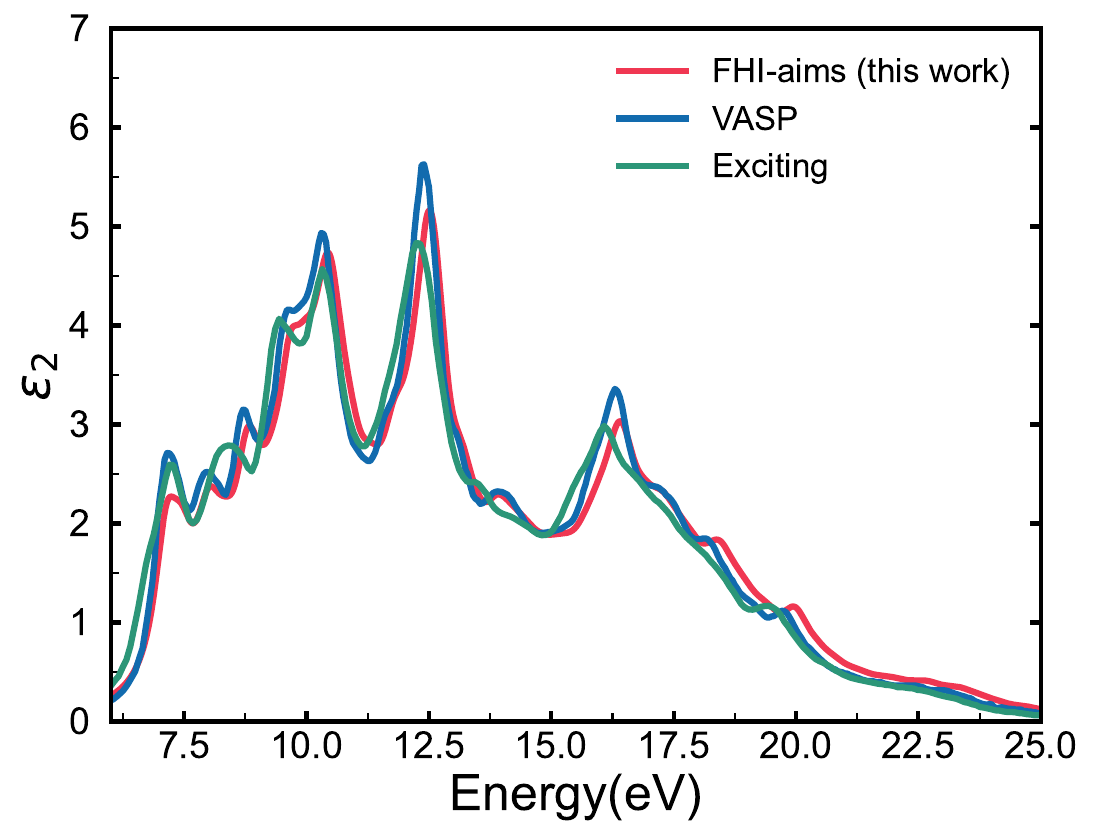}
        \caption{}
        %\caption{Transition Mode 2}
        %\label{fig:MgO-comp}
    \end{subfigure}
    
    \caption{Optical absorption spectrum of MgO from BSE@$G_0W_0$ calculations.
A Lorentzian broadening of $\eta= 0.30 eV$ is used in all cases.
(a) Optical absorption spectrum of MgO with $\Gamma$-centered BZ sampling with dimensions $n \times n \times n$ ($n=11, 13, 15$). Calculations are performed using \textit{tier}  2 NAO basis set complemented by the OBS+4f auxiliary NAO basis set.
(b) Comparison of MgO absorption spectra from our NAO implementation in FHI-aims code with the result from VASP package (PW+PAW) using the $15 \times 15 \times 15$ $\Gamma$-centered BZ sampling. The result from the Exciting code (LAPW+lo) with a $11 \times 11 \times 11$ randomly shifted BZ sampling is also compared. Both the PW+PAW and LAPW+lo results are taken from Ref. \cite{begum2021theoretical}.}
\label{fig:MgO}
\end{figure}

To make a direct comparison, we employ the same computational settings for XC functional, the lattice structure, as well as the BSE active space while BZ sampling varies somewhat for the LAPW+lo calculation reported in Ref.\cite{begum2021theoretical}.  
The DFT-KS calculation is used as the starting point, employing the PBEsol Generalized Gradient Approximation (GGA)\cite{perdew2008restoring, csonka2009assessing}. 
Within this XC functional,
the DFT-optimized equilibrium lattice constant is 4.21 \AA \cite{begum2021theoretical}.
To compute QP energies, a single-shot $G_0W_0$ calculation was performed, utilizing a $7\times7\times7$ $\Gamma$-centered BZ sampling for the dielectric matrix in the self-energy calculation. 
The analytical continuation with the Pade approximation with 100 parameters was performed for $GW$ self-energy on the imaginary axis. 
For the BSE calculation, we employed the $n\times n\times n$ $\Gamma$-centered  BZ sampling with $n=11, 13, 15$. 
In constructing the BSE Hamiltonian,  4 valence bands and 5 conduction bands are included. Here, all calculations were carried out using the \textit{tier} 2 NAO basis set \cite{blum2009ab} and the OBS+4f auxiliary basis set \cite{ihrig2015accurate,ren2021all} for both $G_0W_0$ and BSE calculations. 
In Figure \ref{fig:MgO}(a), we present the optical absorption spectrum with different $n$ for the BZ sampling. 
Similarly to the case of the crystalline silicon, the shape and the relative peak heights of the absorption spectrum are sensitive to the BZ sampling even though the dielectric function is converged
in the $G_0W_0$ calculation already with $7\times7\times7$ $\Gamma$-centered BZ sampling (see Figure S1 in Supplementary Information (SI)). 
As we increase the number of the BZ sampling points, the contribution from high-symmetry k-points decreases for calculating the absorption spectrum. Consequently, we observe the increased splitting of the absorption peaks in the energy range of 8-11 eV. Nevertheless, the absorption spectrum tends to converge as the BZ sampling grid reaches $15\times 15\times 15$, which is the largest grid size used here due to the computational cost and memory, and this is also the most dense BZ sampling reported in the literature \cite{begum2021theoretical}. 

Lastly, we compare the BSE@$GW$ result from our NAO implementation to the results reported in the literature using the planewave-based formulations as shown in Figure \ref{fig:MgO}(b). In particularly, we compare to the results reported using the VASP code and the Exciting code from Ref. \cite{begum2021theoretical}.
The VASP code is based on the PW+PAW scheme\cite{kresse1996efficient,kresse1999ultrasoft}, and the Exciting code is based on LAPW+lo scheme\cite{gulans2014exciting}.
Lorentzian broadening parameter of 0.30 eV was used for the optical absorption spectrum in all calculations. 
The VASP (PP+PAW) result was obtained using the BZ sampling of $15\times15\times15$ centered at the $\Gamma$ point while the Exciting (LAPW+lo) result was obtained with a randomly-shifted $11\times11\times11$ k-grids centered at (0.09, 0.02, 0.04) of the BZ.
%Regarding the first exciton binding energy, we obtained a value of 340 meV, which closely aligns with the values reported in the literature:\rz{ 442 meV (VASP result)\cite{begum2021theoretical} and 429 meV (PW-PP but GW scissor shift result)\cite{fuchs2008efficient}. 
For the lowest excited state, we obtain a large value of 385 meV for the exciton binding energy (i.e. the difference between the excited energy and the quasi-particle enegry gap), which is a good agreement with the reported value of 442 meV using the VASP code  and 435 meV using the exciting code. \cite{begum2021theoretical}.

\begin{figure}
    \centering
        \centering
   \centering
    \begin{subfigure}{0.6\textwidth}
        \centering
        \includegraphics[width=\textwidth]{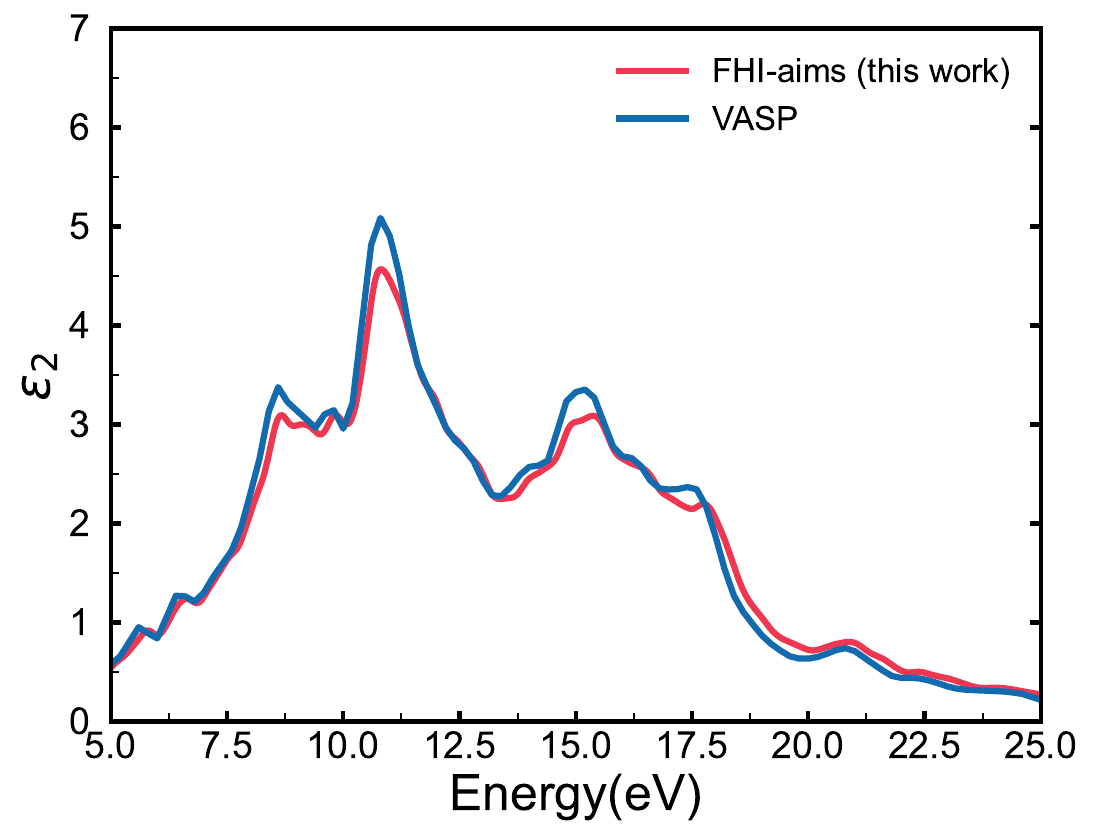}
    \end{subfigure}
   
    \caption{
 Comparison of absorption spectrum of MgO  within independent particle  (IP) approximation from DFT Kohn Sham eigenvalues between FHI-aims calculation (this work) and VASP reference \cite{begum2021theoretical}. FHI-aims calculations are performed under  \textit{tier}  2 NAO basis set complemented by the OBS+4f auxiliary NAO basis set and  $\Gamma$-centered $15 \times 15 \times 15$  BZ sampling. }
 %\vb{VASP reference?}

    \label{fig:MgO-IP}
\end{figure}

Figure \ref{fig:MgO}(b) shows the optical absorption spectrum for our NAO-based implementation, along with the results using the VASP (PP+PAW) and Exciting (LAPW+lo) codes as reported in Ref.\cite{begum2021theoretical}.
The VASP result and the Exciting result agree well above 10 eV but they show noticable differences between 8 eV and 10 eV . 
Using the exactly the same BZ sampling, our NAO-based result and the VASP (PP+PAW) result agree quite well in terms of the peak positions overall. At the same time, a variation between our result and the reported VASP/Exciting results is observed for the amplitudes of some prominent peaks.  To investigate this difference, we examined the optical absorption spectrum within the independent-particle (IP) approximation. Unlike the BSE formalism described in Eq. \ref{eq:absorption_spectrum_eh}, 
the IP absorption spectrum is directly calculated by enumerating all excitation pairs between the valence band and the conduction band of the Kohn-Sham equations,
\begin{equation}\label{eq:absorption_spectrum_noeh}
\epsilon_2(\omega)=\frac{16\pi^2e^2}{\omega^2}\sum_{vc\mathbf{k}}|\mathbf{e}\cdot \langle v\mathbf{k}|\hat{\mathbf{v}}|c\mathbf{k}\rangle |^2\delta(\omega-\epsilon_{c\mathbf{k}}+\epsilon_{v\mathbf{k}}).
\end{equation}
 where $\hat{\mathbf{v}}$ is the velocity operator and $\mathbf{e}$ is the direction of the polarization of light.
As seen in Figure \ref{fig:MgO-IP}, the absorption spectra within the IP approximation using KS states agree well, with some residual discrepancies between our result and the reported result\cite{begum2021theoretical} in the peak intensities, similar to what is seen in the BSE@$GW$ calculation (Figure \ref{fig:MgO}(b)). 

As noted earlier, both the valence states and the low-lying unoccupied states of MgO  using FHI-aims NAO ``\textit{tier} 2'' basis sets are fully converged in KS-DFT calculations \cite{Huhn2017} and, for $G_0W_0$ calculations, are converged to approximately 0.2~eV or better\cite{ren2021all}. These factors would not contribute to the discrepancies seen between our results and those reported in the referenced literature\cite{begum2021theoretical}.
Therefore, the small residual differences in the BSE@$GW$ calculations can be partially ascribed to the differences in the high-lying unoccupied states and the corresponding transition matrix elements on which BSE@$GW$ calculation method relies.

%%%%%%%%%%%%%%%%%%%%%%%%%%%%%%%%%%%%

\section{6. Conclusion}
We described the formulation and algorithms of a new all-electron periodic BSE implementation within the numeric atom-centered orbital (NAO) framework. 
To our knowledge, this is the first all-electron NAO-based BSE implementation that works with periodic boundary conditions. Our implementation was carried out within the FHI-aims code package \cite{blum2009ab,blum2022fhi}. 
We use computed absorption spectra for crystalline silicon (Si) as an example
to demonstrate our implementation and performed systematic convergence tests with respect to the computational parameters including the NAO basis set size, auxiliary
basis set, and Brillouin zone sampling. With the fully-converged result in hand, we make a direct comparison of our
all-electron NAO result to absorption spectra computed using the traditional PW-PP approach implemented in the BerkeleyGW \cite{deslippe2012berkeleygw} code with Quantum Espresso \cite{giannozzi2020quantum} code. 
Having established the excellent agreement with the well-established implementation of BSE@$GW$ based on the PW-PP formalism, 
we also performed calculations on crystalline MgO with a large exciton binding energy.
We compare our NAO-based BSE@$GW$ calculation to available BSE@$GW$ results based on PP+PAW and LAPW+lo formulations from the literature. The calculations show good agreement for the optical absorption spectra, again demonstrating the accuracy of our approach. However, achieving complete convergence of the optical absorption spectrum may also necessitate an exceedingly fine sampling of the BZ as widely discussed in the literature. \cite{sander2015beyond}
Our current implementation supports the standard $\Gamma$-centered sampling.
In future research, we plan to  explore incorporating enhanced BZ sampling techniques for BSE calculation, such as coarse-grained k-grid interpolation \cite{deslippe2012berkeleygw}, Wannier interpolation \cite{kammerlander2012speeding}, and other related methods \cite{sander2015beyond,gillet2016efficient}. As mentioned in the introduction, another important future opportunity is that the groundwork for periodic BSE@$GW$ laid out in the present work should be extendable to much larger systems, by using optimization strategies similar to those that recently enabled accurate periodic exact exchange and hybrid DFT beyond 10,000 atoms using the same underlying NAO-based framework.\cite{Kokott2024}

\section{Acknowledgments}
We thank the Research Computing at the University of North Carolina at Chapel Hill for computer resources.  We thank  Jianhang Xu,  Minye Zhang and Tianyu Zhu for helpful discussion.

\section{Author Contributions}
R.Z. and Y.Y. led and contributed equally to the work. All authors discussed the results and contributed to the final manuscript.
\bibliography{ref}
\end{document}